\documentclass[12pt]{article}

\usepackage{amsmath,amssymb}
\numberwithin{equation}{section}

\evensidemargin \oddsidemargin

\begin{document}

\begin{titlepage}

\renewcommand{\thefootnote}{\fnsymbol{footnote}}

\hfill{hep-th/0303153}

\vspace{15mm}
\baselineskip 9mm
\begin{center}
{\LARGE \bf Type IIA Supergravity Excitations\\
in Plane-Wave Background}
\end{center}

\baselineskip 6mm
\vspace{10mm}
\begin{center}
  O-Kab Kwon$^{a,b}$\footnote{\tt okwon@newton.skku.ac.kr} and
  Hyeonjoon Shin$^a$\footnote{\tt hshin@newton.skku.ac.kr}
  \\[5mm]
  {\sl $^a$BK 21 Physics Research Division and Institute of Basic
    Science \\ Sungkyunkwan University, Suwon 440-746, Korea \\
    $^b$School of Physics, Korea Institute for Advanced Study, Seoul
    130-012, Korea }
\end{center}

\thispagestyle{empty}

\vfill
\begin{center}
{\bf Abstract}
\end{center}
\noindent
We study the low-lying excitations of Type IIA superstring theory in a
plane wave background with 24 supersymmetries.  In the light-cone
gauge, the superstring action has ${\cal N}=(4,4)$ supersymmetry and
is exactly solvable, since it is quadratic in superstring coordinates.
We obtain explicitly the spectrum of the Type IIA supergravity
fluctuation modes in the plane wave background and give its
correspondence with the spectrum of string states from the zero-mode
sector of the light-cone superstring Hamiltonian.

\vspace{20mm}
\end{titlepage}

\baselineskip 6.5mm
\renewcommand{\thefootnote}{\arabic{footnote}}
\setcounter{footnote}{0}

\section{Introduction}

Recently, much attention has been paid to the M theory on the
maximally supersymmetric eleven-dimensional pp-wave background after
the work by Berenstein, Maldacena and Nastase \cite{ber021}.  The
formulation of the theory has been given in the framework of Matrix
model and the resulting Matrix model has turned out to have intriguing
properties \cite{ber021,das185}.  One may take two basic properties,
which may be regarded as the sources of the other.  One is the removal
of flat directions because of the presence of bosonic mass terms and
another is the time dependent supersymmetry making the bosons and
fermions have different masses.  The Matrix model with these
properties shows a peculiar perturbative aspect such as the protected
multiplet and has various kinds of non-perturbative BPS states
\cite{das185}-\cite{par161}.  The discussions on other aspects of the
model may be found in \cite{kim061}-\cite{hyu090}.

In addition to the Matrix model description of the M theory, there is
another possible way of studying the M theory by going down to the ten
dimensional Type IIA superstring theory through the strong-weak
duality between M and Type IIA superstring theory.  In this approach,
there is an advantage that one may use the rather well-developed
machinery of string theory compared to that of Matrix theory.

To obtain the IIA superstring theory from the M theory, one should
pick up an isometry direction on which the M theory is compactified.
As for the eleven-dimensional maximally supersymmetric pp-wave
geometry, there are various spacelike isometries, along which the M
theory can be compactified \cite{mic140}.  One choice of the compact
direction leads to the following ten-dimensional pp-wave background
\cite{ben195,hyu074,sug029}:
\begin{eqnarray}
& & ds^2 = - 2 dx^+ dx^- - A(x^I) (dx^+)^2  + \sum^8_{I=1} (dx^I)^2~,
      \nonumber \\
& & \bar{F}_{+123} = \mu~,~~~ \bar{F}_{+4} = -\frac{\mu}{3}~,
\label{pp-wave}
\end{eqnarray}
where the quantities with bar mean that they are background and we have
defined
\begin{equation}
A(x^I) = \sum^4_{i=1} \frac{\mu^2}{9} (x^i)^2
            +\sum^8_{i'=5} \frac{\mu^2}{36} (x^{i'})^2~.
\end{equation}
It has been shown in \cite{ben195,hyu074} that this background admits
only 24 Killing spinors.

The various aspects of the Type IIA Green-Schwarz (GS) superstring
theory have been studied in \cite{hyu074,sug029,hyu158,hyu343}.  In
particular, in the light-cone gauge formulation, the Type IIA GS
superstring action in this background has been shown to be quadratic
in terms of the string coordinates indicating the exact solvability of
the theory.  Furthermore, the light-cone gauge superstring action has
the interesting linearly-realized worldsheet supersymmetry identified
as ${\cal N}=(4,4)$ \cite{hyu074}.  The situation is similar to the
Type IIB case except that the IIB background is maximally
supersymmetric \cite{met044,met109}.  We note that rather general
discussions on the superstring in the pp-wave background have been
given in \cite{ali037}-\cite{rus179} .

Having the solvable superstring theory, it may be a natural step to
investigate the spectrum of quantized string states and their
dynamics.  In this paper, we concentrate on the low-lying perturbative
string states and study their spectrum in the field theoretic way.
That is, we obtain the physical excitation modes of the Type IIA
supergravity in the pp-wave background, (\ref{pp-wave}), and give
their correspondence with the low-lying string states from the
zero-mode sector of the string theory.  As alluded above, the study of
IIA string theory is motivated by a hope to understand the M theory in
a controllable way.  In the situation that we have the Matrix model as
another way of studying the M theory, it may be expected that the
low-energy perturbative study of the IIA string theory is helpful in
uncovering the physics related to the perturbative spectrum of the
Matrix model in the pp-wave.  The work in this paper may be regarded
as the first step in this direction of study.  We note that there have
been other related works for the supergravity spectrum
\cite{fer175,bre308}.

The organization of this paper is as follows.  In section
\ref{string}, following Refs.~\cite{hyu074,hyu158}, we review the
derivation of Type IIA GS superstring action in the pp-wave
background, (\ref{pp-wave}), which is given in the light-cone gauge
formulation, and then the quantization of the superstring.  After the
review, we give the low-lying spectrum of string states.  In section
\ref{sugra}, we consider the fields of Type IIA supergravity in the
pp-wave background and obtain the physical supergravity excitation
modes around the background.  We shall see how the spectrum of the
supergravity modes corresponds to that of low-lying string states
obtained in section \ref{string}.  Finally, the conclusion and
discussion follow in section \ref{concl}.

\section{Type IIA superstring in plane-wave background}
\label{string}

In this section, we review the light-cone gauge fixed Type IIA GS
superstring action in the pp-wave background and its quantization
following \cite{hyu074,hyu158}.

It is very complicate to get the full expression of the GS superstring
action in the general background (see, for example,
\cite{cve202,hyu247}).  However, in the case at hand, we can use the
fact that eleven-dimensional pp-wave geometry can be thought as a
special limit of $AdS_4 \times S^7$ geometry on which the full
supermembrane action is constructed using coset method \cite{dew209}.
The full IIA GS superstring action on this geometry can be obtained by
the double dimensional reduction \cite{duf70} of the supermembrane
action of \cite{dew209} in the Penrose limit \cite{pen271}.  The
superstring action is simplified drastically in the light-cone gauge
chosen as
\begin{equation}
\Gamma^+ \theta = 0 ~,~~~
X^+ = \alpha' p^+  \tau ~,
\label{kfix}
\end{equation}
where $p^+$ is the total momentum conjugate to $X^-$ and $\tau$ is the
worldsheet time coordinate.  In this light-cone gauge, IIA string
action is given by\footnote{$\eta^{mn}$ is the flat worldsheet metric
  with $m,n$ taking values of $\tau,\sigma$.}
\begin{eqnarray}
S_{LC}
 &=& - \frac{1}{4 \pi \alpha'} \int  d^2 \sigma
 \Bigg[ \eta^{mn} \partial_m X^I \partial_n X^I
      + \frac{m^2}{9} (X^i)^2
      + \frac{m^2}{36} (X^{i'})^2
                       \nonumber \\
 & & + \bar{\theta} \Gamma^- \partial_\tau \theta
     + \bar{\theta} \Gamma^{-9} \partial_\sigma \theta
     - \frac{m}{4} \bar{\theta} \Gamma^-
        \left( \Gamma^{123} + \frac{1}{3} \Gamma^{49} \right)
        \theta
  \Bigg] ~,
\end{eqnarray}
where
\begin{equation}
m\equiv \mu \alpha' p^+
\end{equation}
is a mass parameter which characterizes the masses of the worldsheet
fields, and the Majorana fermion $\theta$ is the combination of
Majorana-Weyl fermions $\theta^1$ and $\theta^2$ with opposite ten
dimensional $SO(1,9)$ chiralities, that is, $\theta = \theta^1 +
\theta^2$. (1 (2) is for positive (negative) chirality.)  Therefore
the light-cone gauge-fixed action $S_{LC}$ is quadratic in bosonic as
well as fermionic fields and thus describes a free theory much the
same as in the IIB string theory \cite{met044} on the pp-wave geometry
\cite{bla242}.

The characteristic feature of IIA string theory in pp-wave background
is the structure of worldsheet supersymmetry.  Sixteen spacetime
supersymmetries with transformation parameter $\epsilon$ satisfying
$\Gamma^+ \epsilon = 0$ are non-linearly realized on the worldsheet
action.  As is typical in light-cone GS superstring, the remaining
eight spacetime supersymmetries, combined with appropriate kappa
transformations, turn into worldsheet (4,4) supersymmetry of
Yang-Mills type \cite{hyu074}.  In order to see this more clearly, we
rewrite the action $S_{LC}$ in the 16 component spinor notation.  We
should first introduce the representation for $SO(1,9)$ gamma matrices
which we take as
\[
 \Gamma^0= -i\sigma^2 \otimes {\bf 1}_{16}~,~~~
 \Gamma^{11}= \sigma^1 \otimes {\bf 1}_{16}~,~~~
 \Gamma^I= \sigma^3 \otimes \gamma^I~,
\]
\begin{equation}
 \Gamma^9= -\sigma^3 \otimes \gamma^9~,~~~
 \Gamma^{\pm} = \frac{1}{\sqrt{2}}(\Gamma^0 \pm \Gamma^{11})~,
\label{gamma}
\end{equation}
where $\sigma$'s are Pauli matrices, and ${\bf 1}_{16}$ the $16 \times
16$ unit matrix. $\gamma^I$ are the $16 \times 16$ symmetric real
gamma matrices satisfying the spin$(8)$ Clifford algebra $\{ \gamma^I,
\gamma^J \} = 2 \delta^{IJ}$, which are reducible to the ${\bf
  8_s}+{\bf 8_c}$ representation of spin$(8)$.  We note that, since
the pp-wave background (\ref{pp-wave}) has been obtained by
compactifying the eleven dimensional pp-wave along the $x^9$ direction
\cite{hyu074}, $\Gamma^9$ is the $SO(1,9)$ chirality operator and
$\gamma^9$ becomes $SO(8)$ chirality operator,
\begin{equation}
\gamma^9 = \gamma^1 \cdots \gamma^8~.
\end{equation}
Then, with the spinor notation $\theta^A = \frac{1}{2^{1/4}} \left(
\begin{array}{c} 0 \\ \psi^A
\end{array} \right)$ satisfying the light-cone gauge (\ref{kfix})
(Superscript $A$ denotes the $S(1,9)$ chirality.), the action $S_{LC}$
becomes
\begin{eqnarray}
S_{LC}
 &=&  - \frac{1}{4 \pi \alpha'} \int  d^2 \sigma
 \Bigg[ \eta^{mn} \partial_m X^I \partial_n X^I
      + \frac{m^2}{9} (X^i)^2
      + \frac{m^2}{36} (X^{i'})^2
                       \nonumber \\
 & & - i \psi_+^1 \partial_+  \psi^1_+
     - i \psi_-^1 \partial_+  \psi^1_-
     - i \psi^2_+ \partial_- \psi^2_+
     - i \psi^2_- \partial_- \psi^2_-
     +2i \frac{m}{3} \psi^2_+ \gamma^4 \psi^1_-
     - 2i \frac{m}{6} \psi^2_- \gamma^4 \psi^1_+
 \Bigg]~,
                       \nonumber \\
\label{lc-action}
\end{eqnarray}
where $\partial_\pm =\partial_\tau\pm \partial_\sigma$.  Here the sign
of subscript in $\psi^A_\pm$ represents the eigenvalue of
$\gamma^{1234}$.  In our convention, fermion has the same $SO(1,9)$
and $SO(8)$ chirality measured by $\Gamma^9$ and $\gamma^9$,
respectively.

Thus, among sixteen fermionic components in total, eight with
$\gamma^{12349}=1$ have the mass of $m/6$ and the other eight with
$\gamma^{12349}=-1$ the mass of $m/3$, which are identical with the
masses of bosons. Therefore the theory contains two supermultiplets
$(X^i, \psi^1_-, \psi^2_+)$ and $(X^{i'}, \psi^1_+, \psi^2_-)$ of
worldsheet (4,4) supersymmetry with the masses $m/3$ and $m/6$,
respectively.

Let us now turn to the quantization of closed string in the pp-wave
background \cite{hyu158} and consider the low-lying string states
constructed by acting the zero-mode creation operators on the vacuum
which should correspond to the Type IIA supergravity excitations in
the pp-wave background.

We first consider the bosonic sector of the theory.  The equations of
motion for the bosonic coordinates $X^I$ are read off from the action
(\ref{lc-action}) as
\begin{equation}
\eta^{mn} \partial_m \partial_n X^i
  - \left( \frac{m}{3} \right)^2 X^i = 0 ~,~~~
\eta^{mn} \partial_m \partial_n X^{i'}
  - \left( \frac{m}{6} \right)^2 X^{i'} = 0~,
\label{beom}
\end{equation}
where the fields are subject to the periodic boundary condition,
\begin{equation}
X^I (\tau, \sigma+2 \pi) = X^I (\tau, \sigma)~.
\end{equation}
The solutions of the above equations are given in the form of mode
expansion and found to be
\begin{eqnarray}
X^i (\tau,\sigma)
  &=& x^i \cos \left( \frac{m}{3} \tau \right)
     + \alpha' p^i \frac{3}{m}
             \sin \left( \frac{m}{3} \tau \right)
   + i \sqrt{ \frac{\alpha'}{2} } \sum_{n \neq 0}
         \frac{1}{\omega_n}
      ( \alpha^i_n \phi_n (\tau, \sigma )
    + \tilde{\alpha}^i_n \tilde{\phi}_n (\tau, \sigma) ) ~,
                                   \nonumber \\
X^{i'} (\tau,\sigma)
  &=& x^{i'} \cos \left( \frac{m}{6} \tau \right)
     + \alpha' p^{i'} \frac{6}{m}
              \sin \left( \frac{m}{6} \tau \right)
   + i \sqrt{ \frac{\alpha'}{2} } \sum_{n \neq 0}
         \frac{1}{\omega'_n}
      ( \alpha^{i'}_n \phi'_n (\tau, \sigma)
     + \tilde{\alpha}^{i'}_n \tilde{\phi}'_n (\tau, \sigma) )~,
                                   \nonumber \\
\label{bmode}
\end{eqnarray}
where $x^I$ and $p^I$ are center-of-mass variables defined in the
usual manner, coefficients for the zero-modes, and $\alpha^I_n$ and
$\tilde{\alpha}^I_n$ are the expansion coefficients for the non-zero
modes.  The basis functions for the non-zero modes are given by
\begin{eqnarray}
& & \phi_n (\tau, \sigma) = e^{-i \omega_n \tau - i n \sigma}~,~~~
\tilde{\phi}_n (\tau, \sigma) = e^{-i \omega_n \tau + i n \sigma}~,
\label{mode}  \\
& & \phi'_n (\tau, \sigma) = e^{-i \omega'_n \tau - i n \sigma}~,~~~
\tilde{\phi}'_n (\tau, \sigma)
  = e^{-i \omega'_n \tau + i n \sigma}~,
\label{pmode}
\end{eqnarray}
with the wave frequencies
\begin{equation}
\omega_n = {\rm sign}(n)
      \sqrt{ \left( \frac{m}{3} \right)^2 + n^2 }~,~~~
\omega'_n = {\rm sign}(n)
      \sqrt{ \left( \frac{m}{6} \right)^2 + n^2 }~.
\end{equation}
We note that the reality of $X^I$ requires that $\alpha^{I \dagger}_n
= \alpha^I_{-n}$ and $\tilde{\alpha}^{I \dagger}_n =
\tilde{\alpha}^I_{-n}$.

We promote the expansion coefficients in the mode expansions
(\ref{bmode}) to operators.  By using the canonical equal time
commutation relations for the bosonic fields,
\begin{equation}
[ X^I (\tau, \sigma), {\cal P}^J (\tau, \sigma') ]
= i \delta^{IJ} \delta (\sigma - \sigma') ~,
\label{bcom}
\end{equation}
where ${\cal P}^J = \partial_\tau X^J / 2 \pi \alpha'$ is the
canonical conjugate momentum of $X^J$, we have the following
commutation relations between mode operators:
\begin{equation}
[ x^I, p^J ] = i \delta^{IJ} ~,~~~
[ \alpha^i_n, \alpha^j_m ] = \omega_n \delta^{ij} \delta_{n+m,0}~,~~~
[ \alpha^{i'}_n, \alpha^{j'}_m ] =
   \omega'_n \delta^{i' j'} \delta_{n+m,0}~.
\label{bmcom}
\end{equation}

Let us next turn to the fermionic sector of the theory. The fermionic
fields are split into two parts according to the $(4,4)$
supersymmetry; $(\psi^1_-, \psi^2_+)$ and $(\psi^1_+, \psi^2_-)$.  We
first consider the former case.  The equations of motion for
$\psi^1_-$ and $\psi^2_+$ are obtained as
\begin{equation}
\partial_+ \psi^1_- + \frac{m}{3} \gamma^4 \psi^2_+ = 0 ~,~~~
\partial_- \psi^2_+ - \frac{m}{3} \gamma^4 \psi^1_- = 0 ~.
\label{feom}
\end{equation}
The non-zero mode solutions of these equations are given by using the
modes, (\ref{mode}). For the zero mode part of the solution, we impose
a condition that, at $\tau=0$, the solution behaves just as that of
massless case.  The mode expansions for the fermionic coordinates are
then
\begin{eqnarray}
\psi^1_- (\tau, \sigma)
 &=& c_0 \tilde{\psi}_0 \cos \left( \frac{m}{3} \tau \right)
     - c_0 \gamma^4 \psi_0 \sin \left( \frac{m}{3} \tau \right)
                              \nonumber \\
 & & + \sum_{n \neq 0} c_n
      \left(
         \tilde{\psi}_n \tilde{\phi}_n (\tau, \sigma)
         - i \frac{3}{m} (\omega_n - n)
            \gamma^4 \psi_n \phi_n (\tau, \sigma )
      \right)~,
                               \nonumber \\
\psi^2_+ (\tau, \sigma)
 &=& c_0 \psi_0 \cos \left( \frac{m}{3} \tau \right)
     + c_0 \gamma^4 \tilde{\psi}_0
        \sin \left( \frac{m}{3} \tau \right)
                              \nonumber \\
 & & + \sum_{n \neq 0} c_n
      \left(
         \psi_n \phi_n (\tau, \sigma)
         + i \frac{3}{m} (\omega_n - n)
            \gamma^4 \tilde{\psi}_n \tilde{\phi}_n (\tau, \sigma )
      \right)~,
\label{fmode}
\end{eqnarray}
where $\gamma^{1234} \psi_n = \psi_n$ and $\gamma^{1234}
\tilde{\psi}_n = -\tilde{\psi}_n$ for all $n$, and $c_n$ are the
normalization constants given by
\[
c_0 = \sqrt{\alpha'}~,~~~
c_n = \frac{\sqrt{\alpha'}}{\sqrt{1 +
          \left( \frac{3}{m} \right)^2 (\omega_n - n)^2 } }~.
\]
Promoting the expansion coefficients to operators and using the
canonical equal time anti-commutation relations,
\begin{equation}
\{ \psi^A_\pm (\tau, \sigma), \psi^B_\pm (\tau, \sigma') \}
= 2 \pi \alpha' \delta^{AB} \delta (\sigma - \sigma') ~,
\label{fcom}
\end{equation}
the following anti-commutation relations between mode operators are
obtained.
\begin{equation}
\{ \psi_n , \psi_m \} = \delta_{n+m,0} ~,~~~
\{ \tilde{\psi}_n, \tilde{\psi}_m \} = \delta_{n+m,0}~.
\label{fmcom}
\end{equation}

The quantization of fermionic coordinates in the other $(4,4)$
supermultiplet proceed along the same way.  The equations of motion
for $\psi^1_+$ and $\psi^2_-$ are respectively
\begin{equation}
\partial_+ \psi^1_+ - \frac{m}{6} \gamma^4 \psi^2_- = 0 ~,~~~
\partial_- \psi^2_- + \frac{m}{6} \gamma^4 \psi^1_+ = 0 ~,
\label{fpeom}
\end{equation}
whose solutions are found to be
\begin{eqnarray}
\psi^1_+ (\tau, \sigma)
 &=& c'_0 \tilde{\psi}'_0 \cos \left( \frac{m}{6} \tau \right)
     + c'_0 \gamma^4 \psi'_0 \sin \left( \frac{m}{6} \tau \right)
                              \nonumber \\
 & & + \sum_{n \neq 0} c'_n
      \left(
         \tilde{\psi}'_n \tilde{\phi}'_n (\tau, \sigma)
         + i \frac{6}{m} (\omega'_n - n)
            \gamma^4 \psi'_n \phi'_n (\tau, \sigma )
      \right)~,
                               \nonumber \\
\psi^2_- (\tau, \sigma)
 &=& c'_0 \psi'_0 \cos \left( \frac{m}{6} \tau \right)
     - c'_0 \gamma^4 \tilde{\psi}'_0
        \sin \left( \frac{m}{6} \tau \right)
                              \nonumber \\
 & & + \sum_{n \neq 0} c'_n
      \left(
         \psi'_n \phi'_n (\tau, \sigma)
         - i \frac{6}{m} (\omega'_n - n)
            \gamma^4 \tilde{\psi}'_n \tilde{\phi}'_n (\tau, \sigma )
      \right)~,
\label{fpmode}
\end{eqnarray}
where $\gamma^{1234} \psi'_n = -\psi'_n$, $\gamma^{1234}
\tilde{\psi}'_n = \tilde{\psi}'_n$, and
\[
c'_0 = \sqrt{\alpha'}~,~~~
c'_n = \frac{\sqrt{\alpha'}}{\sqrt{1 +
          \left( \frac{6}{m} \right)^2 (\omega'_n - n)^2 } }~.
\]
Then the equal time anti-commutation relations, (\ref{fcom}), lead us
to have
\begin{equation}
\{ \psi'_n , \psi'_m \} = \delta_{n+m,0} ~,~~~
\{ \tilde{\psi}'_n, \tilde{\psi}'_m \} = \delta_{n+m,0}~.
\label{fpmcom}
\end{equation}

We now consider the light-cone Hamiltonian of the theory, which is
written as\footnote{ From now on, we set $2 \pi
  \alpha' = 1$ for notational convenience.}
\begin{equation}
H_{LC}
 = \int^{2\pi}_0 d \sigma P^-
 = \frac{2 \pi}{p^+} \int^{2\pi}_0 d \sigma {\cal H} ~.
\label{lc-ham}
\end{equation}
The ${\cal H}$ is the Hamiltonian density obtained from
Eq.~(\ref{lc-action}) as
\begin{eqnarray}
{\cal H}
 &=&  \frac{1}{2} ({\cal P}^I)^2
    + \frac{1}{2} (\partial_\sigma X^I)^2
    +\frac{1}{2} \left( \frac{m}{3} \right)^2 (X^i)^2
    +\frac{1}{2} \left( \frac{m}{6} \right)^2 (X^{i'})^2
  \nonumber \\
 & & - \frac{i}{2} \psi^1_- \partial_\sigma \psi^1_-
     + \frac{i}{2} \psi^2_+ \partial_\sigma \psi^2_+
     + i \frac{m}{3} \psi^2_+ \gamma^4 \psi^1_-
  \nonumber \\
 & & - \frac{i}{2} \psi^1_+ \partial_\sigma \psi^1_+
     + \frac{i}{2} \psi^2_- \partial_\sigma \psi^2_-
     - i \frac{m}{6} \psi^2_- \gamma^4 \psi^1_+ ~.
\end{eqnarray}
By plugging the mode expansions for the fields, Eqs. (\ref{bmode}),
(\ref{fmode}), and (\ref{fpmode}), into Eq. (\ref{lc-ham}), the
light-cone Hamiltonian becomes
\begin{equation}
H_{LC} = E_0 + E + \tilde{E}~,
\label{lc-h}
\end{equation}
where $E_0$ is the zero mode contribution and $E$, $\tilde{E}$ are the
contributions of the non-zero modes:
\begin{eqnarray}
E_0 &=& \frac{2\pi^2}{p^+}
  \left( \left( \frac{p^I}{2\pi} \right)^2
       + \left( \frac{m}{3} \right)^2 (x^i)^2
       + \left( \frac{m}{6} \right)^2 (x^{i'})^2
       - \frac{i}{\pi} \frac{m}{3} \tilde{\psi}_0 \gamma^4 \psi_0
       + \frac{i}{\pi} \frac{m}{6} \tilde{\psi}'_0 \gamma^4 \psi'_0
  \right)~,
  \nonumber \\
E &=& \frac{\pi}{p^+} \sum_{n \neq 0}
  ( \alpha^I_{-n} \alpha^I_n
   + \omega_n \psi_{-n} \psi_n
   + \omega'_n \psi'_{-n} \psi'_n
  )~,
  \nonumber \\
\tilde{E}
  &=& \frac{\pi}{p^+} \sum_{n \neq 0}
  ( \tilde{\alpha}^I_{-n} \tilde{\alpha}^I_n
   + \omega_n \tilde{\psi}_{-n} \tilde{\psi}_n
   + \omega'_n \tilde{\psi}'_{-n} \tilde{\psi}'_n
  )~.
\end{eqnarray}

In the quantized version, the modes in the expression of Hamiltonian
become operators with the commutation relations, (\ref{bmcom}),
(\ref{fmcom}), and (\ref{fpmcom}), and should be properly normal
ordered.  For the string oscillator contributions, $E$ and
$\tilde{E}$, we place operator with negative mode number to the left
of operator with positive mode number as in the flat case.  The normal
ordered expressions of them are then given by
\begin{eqnarray}
E &=& \frac{2 \pi}{p^+} \sum^\infty_{n =1}
  ( \alpha^I_{-n} \alpha^I_n
   + \omega_n \psi_{-n} \psi_n
   + \omega'_n \psi'_{-n} \psi'_n
  )~,
  \nonumber \\
\tilde{E}
  &=& \frac{2 \pi}{p^+} \sum^\infty_{n =1}
  ( \tilde{\alpha}^I_{-n} \tilde{\alpha}^I_n
   + \omega_n \tilde{\psi}_{-n} \tilde{\psi}_n
   + \omega'_n \tilde{\psi}'_{-n} \tilde{\psi}'_n
  )~.
\label{nzero-h}
\end{eqnarray}
Here we note that there is no zero-point energy because bosonic
contributions are exactly canceled by those of fermions.

The zero mode contribution is the Hamiltonian for the simple harmonic
oscillators and massive fermions.  For the bosonic part, we introduce
the usual creation and annihilation operators as
\begin{eqnarray}
& & a^{i \dagger} = \sqrt{ \frac{3 \pi}{m} }
      \left( \frac{p^i}{2 \pi} + i \frac{m}{3} x^i \right)~,
 ~~~a^i = \sqrt{ \frac{3 \pi}{m} }
      \left( \frac{p^i}{2 \pi} - i \frac{m}{3} x^i \right)~,
  \nonumber \\
& & a^{i' \dagger} = \sqrt{ \frac{6 \pi}{m} }
    \left( \frac{p^{i'}}{2 \pi} + i \frac{m}{6} x^{i'} \right)~,
 ~~~a^{i'} = \sqrt{ \frac{6 \pi}{m} }
    \left( \frac{p^{i'}}{2 \pi} - i \frac{m}{6} x^{i'} \right)~,
\label{bz}
\end{eqnarray}
whose commutation relations are read as, from Eq.~(\ref{bmcom}),
\begin{equation}
[ a^I, a^{J \dagger} ] = \delta^{IJ}~.
\end{equation}
As for the fermionic creation and annihilation operators, we take the
following combination of modes.
\begin{eqnarray}
& & \chi^\dagger = \frac{1}{\sqrt{2}}
             ( \psi_0 - i \gamma^4 \tilde{\psi}_0 )~,
 ~~~\chi = \frac{1}{\sqrt{2}}
             ( \psi_0 + i \gamma^4 \tilde{\psi}_0 )~,
  \nonumber \\
& & \chi'^\dagger = \frac{1}{\sqrt{2}}
             ( \psi'_0 + i \gamma^4 \tilde{\psi}'_0 )~,
 ~~~\chi' = \frac{1}{\sqrt{2}}
             ( \psi'_0 - i \gamma^4 \tilde{\psi}'_0 )~,
\label{fz}
\end{eqnarray}
where $\gamma^{12349} \chi = - \chi$ and $\gamma^{12349} \chi' =
\chi'$.  From Eqs.~(\ref{fmcom}) and (\ref{fpmcom}), the
anti-commutation relations between these operators become
\begin{equation}
\{ \chi, \chi^\dagger \} = 1 ~, ~~~
\{ \chi', \chi'^\dagger \} = 1 ~.
\end{equation}
In terms of the operators introduced above, Eqs.~(\ref{bz}) and
(\ref{fz}), the normal ordered zero mode contribution to the
light-cone Hamiltonian is then given by
\begin{equation}
E_0 = \frac{\mu}{6}
     ( 2 a^{i \dagger} a^i + a^{i' \dagger} a^{i'}
       + 2 \chi^\dagger \chi + \chi'^\dagger \chi' )~,
\label{zero-h}
\end{equation}
which has vanishing zero-point energy as in the case
of string oscillator contributions.

Though the string physics is described by the light-cone Hamiltonian,
(\ref{lc-h}), there is a constraint constraining the string states,
which is the usual Virasoro constraint imposing the invariance under
the translation in $\sigma$ direction.  In the light-cone gauge, the
Virasoro constraint is given by
\begin{equation}
\int^{2\pi}_0 d \sigma
 \left( - \frac{1}{2\pi} p^+ \partial_\sigma X^-
   + {\cal P}^I \partial_\sigma X^I
   + \frac{i}{2} \psi^A_+ \partial_\sigma \psi^A_+
   + \frac{i}{2} \psi^A_- \partial_\sigma \psi^A_-
 \right) = 0 ~.
\end{equation}
The integration of the first integrand vanishes trivially since $p^+$
is constant, and the remaining parts give us the following constraint.
\begin{equation}
N = \tilde{N} ~,
\label{vira}
\end{equation}
where $N$ and $\tilde{N}$ are defined as
\begin{eqnarray}
N &=& \sum^\infty_{n=1} n
   \left( \frac{1}{\omega_n} \alpha^i_{-n} \alpha^i_n
         +\frac{1}{\omega'_n} \alpha^{i'}_{-n} \alpha^{i'}_n
         + \psi_{-n} \psi_n + \psi'_{-n} \psi'_n
   \right) ~,
  \nonumber \\
\tilde{N}
  &=&\sum^\infty_{n=1} n
   \left( \frac{1}{\omega_n} \tilde{\alpha}^i_{-n}
                             \tilde{\alpha}^i_n
         +\frac{1}{\omega'_n} \tilde{\alpha}^{i'}_{-n}
                              \tilde{\alpha}^{i'}_n
         + \tilde{\psi}_{-n} \tilde{\psi}_n
         + \tilde{\psi}'_{-n} \tilde{\psi}'_n
   \right) ~.
\end{eqnarray}

The normal ordered expressions Eqs.~(\ref{nzero-h}) and (\ref{zero-h})
now constitute the quantum light-cone Hamiltonian, which implicitly
defines the vacuum $|0 \rangle$ of the quantized theory as a state
annihilated by string oscillation operators with positive mode number,
that is $n \ge 1$, and zero mode operators $a^I$, $\chi$, and $\chi'$
defined in Eqs.~(\ref{bz}) and (\ref{fz}).  Actually, the vacuum
defined in this paper, especially the vacuum state in the zero mode
sector, is not unique but one of the possible Clifford vacua, since
our theory is massive and there can be various definitions for the
creation and annihilation operators.  This is also the case for the
IIB superstring in pp-wave background and has been discussed in
\cite{met109}.  However, considering the regularity of states at $\tau
\rightarrow i \infty$ that has been pointed out in \cite{rus179}, our
definition is a natural one.

The low-lying string states are obtained by acting the fermionic and
bosonic zero-mode creation operators on the vacuum $| 0 \rangle$ and
correspond to the excitation modes of Type IIA supergravity fields
expanded near the plane-wave background.  Among the string states,
those constructed by using only the fermionic zero-modes, that is,
$(\chi^\dagger)^n (\chi'^\dagger)^{n'} |0 \rangle $ with $n,n' =
0,1,2,3,4$, correspond to the supergravity excitation modes with the
minimal light-cone energies, and the string states obtained by acting
the bosonic zero-modes on them are related to the supergravity modes
with higher light-cone energies.  In table \ref{string-states}, we
list the states $(\chi^\dagger)^n (\chi'^\dagger)^{n'} |0 \rangle $
according to their light-cone energy $H_{LC}$, which is simply $E_0$
of Eq.~(\ref{zero-h}), in units of $\mu / 6$.  We shall see in the
next section how these string states correspond to the supergravity
excitation modes.

\begin{table}
\begin{center}
\begin{tabular}{|c|lr|lr|}
\hline
$e_0$ &   Bosonic states & $N_B$ & Fermionic states & $N_F$ \\
\hline
0   & (0,0)            & 1       &                  & 0     \\
1   &                  & 0       & (0,1)            & 4     \\
2   & (0,2)            & 6       & (1,0)            & 4     \\
3   & (1,1)            & 16      & (0,3)            & 4     \\
4   & (2,0) (0,4)      & 7       & (1,2)            & 24    \\
5   & (1,3)            & 16      & (2,1)            & 24    \\
6   & (2,2)            & 36      & (3,0) (1,4)      & 8     \\
7   & (3,1)            & 16      & (2,3)            & 24    \\
8   & (4,0) (2,4)      & 7       & (3,2)            & 24    \\
9   & (3,3)            & 16      & (4,1)            & 4     \\
10  & (4,2)            & 6       & (3,4)            & 4     \\
11  &                  & 0       & (4,3)            & 4     \\
12  & (4,4)            & 1       &                  & 0     \\
\hline
    &                  & 128     &                  & 128   \\
\hline
\end{tabular}
\end{center}
\caption{Low-lying string states constructed by acting fermionic
zero-mode operators on the vacuum; $(\chi^\dagger)^n
(\chi'^\dagger)^{n'} |0 \rangle $ with $n,n' = 0,1,2,3,4$. The
states are characterized by $(n,n')$. $N_B(N_F)$ is the
number of bosonic (fermionic) degrees of freedom. $e_0$ is
the light-cone energy in units of $\mu / 6$, that is,
$E_0 = \mu e_0 /6$.}\label{string-states}
\end{table}

\section{Supergravity excitation spectrum}
\label{sugra}

The equations of motion for the Type IIA supergravity fields expanded
to linear order in fluctuations on the plane-wave background can be
used to determine the light-cone energy spectrum of the fluctuating
fields. The equations for fluctuations in the plane-wave
background~(\ref{pp-wave}) has the following typical form
\begin{equation}\label{seqn}
\left(\Box + i \alpha \partial_-\right)\varphi = 0
\end{equation}
with
\begin{eqnarray}\nonumber
\Box \equiv \frac{1}{\sqrt{-\bar{g}}}
\partial_\mu \left(\sqrt{-\bar{g}}
    \bar{g}^{\mu\nu}\partial_\nu  \right)
= -2 \partial_+\partial_- + A(x^I) \partial_-^2 + \partial_I^2,
\end{eqnarray}
where $\bar{g} \equiv \det \bar{g}_{\mu\nu}= -1$ and $\alpha$ is an
arbitrary constant.  In the light-cone description where $x^+$ is the
evolution parameter of the system, the fluctuation field $\varphi$ can
be expressed by using the Fourier transformation as follows
\begin{equation}
\varphi(x^+, x^-, x^I) = \int \frac{dp^+ d^8 p^I}{ ( 2\pi)^{9/2}}
e^{i ( - x^- p^+ + x^I p^I)} \tilde \varphi ( x^+, p^+, p^I),
\end{equation}
where $\tilde \varphi$ satisfies
\begin{equation} \label{LEE}
\left[ 2 p^+ P^- + (p^+)^2 \left(\sum_{i = 1}^{4}\frac{\mu^2}{9}
\partial_{p^i}^2  + \sum_{i'=5}^8 \frac{\mu^2}{36} \partial_{p^{i'}}^2
\right) - p_I^2 + \alpha p^+\right] \varphi = 0.
\end{equation}
From the Eq.~(\ref{LEE}), we obtain the light-cone Hamiltonian,
\begin{equation}\label{LE}
H \equiv i\partial_+ = P^-
       = \frac1{2p^+} \left( (p^I)^2 - m_1^2
         \sum_{i =1}^4 \partial_{p^i}^2
         - m_2^2 \sum_{i'= 5}^8\partial_{p^{i'}}^2 \right)
         - \frac{\alpha}{2},
\end{equation}
where $ m_1 = \frac{\mu}3 p^+$, $ m_2 = \frac{\mu}6 p^+$.  To obtain
the light-cone energy spectrum, we introduce two sets of creation and
annihilation operators,
\begin{eqnarray}
&& a^{i \dagger} \equiv \frac1{\sqrt{2m_1}}
    ( p^i - m_1 \partial_{p^i}),
\quad
       ~~a^i \equiv \frac1{\sqrt{2m_1}} ( p^i + m_1 \partial_{p^i}),
\quad
       ~~[a^i, {a^\dagger}^j] = \delta^{ij}, \\
&& a^{i' \dagger} \equiv \frac1{\sqrt{2m_2}}
   ( p^{i'} - m_2 \partial_{p^{i'}}),
\quad
        a^{i'} \equiv \frac1{\sqrt{2m_2}} ( p^{i'} + m_2 \partial_{p^{i'}}),
\quad
        [a^{i'}, {a^\dagger}^{j'}] = \delta^{i'j'} ~.
\end{eqnarray}
Then the normal ordered expression of the light-cone Hamiltonian is
given by
\begin{equation}\label{norH}
H = \frac{\mu}3 \sum_{i=1}^4 a^{i \dagger} a^i
    + \frac{\mu}6 \sum_{i'=1}^4 a^{i' \dagger} a^{i'}
    + \mu - \frac{\alpha}{2}.
\end{equation}
From this relation (\ref{norH}), we see that the fluctuation field which
satisfies the Eq.~(\ref{seqn}) has the minimal light-cone energy
${\cal E}_0$ defined as
\begin{equation}\label{lstE}
\frac{\mu}{6} {\cal E}_0 = \mu - \frac{\alpha}2 ~,
\end{equation}
which will be used to characterize the excitation modes of the IIA
supergravity in the pp-wave background.

\subsection{Bosonic excitations}
\label{b-sugra}

In the bosonic sector of the Type IIA supergravity, we have five
fields, which are dilaton $\Phi$, graviton $g_{\mu\nu}$, NS-NS
two-form gauge field $B_{\mu\nu}$, and two R-R gauge fields $A_\mu$
and $A_{\mu\nu\rho}$.  The equations of motion for these bosonic
fields are, in the Einstein frame,
\begin{eqnarray}
&& \nabla^2 \Phi =
         - \frac1{12} e^{-\Phi} H_{\mu\nu\rho} H^{\mu\nu\rho}
         + \frac38 e^{\frac32 \Phi} F_{\mu\nu} F^{\mu\nu}
         + \frac1{96} e^{\frac12\Phi}  \tilde{F}_{\mu\nu\rho\sigma}
                                     \tilde{F}^{\mu\nu\rho\sigma} ~,
\label{EQ1} \\
&& \nabla_\mu \left( e^{\frac32 \Phi} F^{\mu\nu} \right) =
         - \frac16 e^{\frac12 \Phi} H_{\mu\rho\sigma}
                           \tilde{F}^{\mu\nu\rho\sigma} ~,
\label{EQ2} \\
&& \nabla_\mu \left(  e^{-\Phi} H^{\mu\nu\rho}
                    + e^{\frac{1}{2} \Phi} A_\sigma
                           \tilde{F}^{\sigma\mu\nu\rho}
              \right) =
   \frac{1}{2\cdot (4!)^2}
   \frac{\epsilon^{\nu\rho \mu_1\cdots \mu_8}}{\sqrt{-g}}
              F_{\mu_1 \cdots \mu_4} F_{\mu_5 \cdots \mu_8} ~,
\label{EQ3} \\
&& \nabla_\mu \left( e^{\frac12 \Phi} \tilde{F}^{\mu\nu\rho\sigma}
              \right) =
      -\frac1{6\cdot 4!}
       \frac{\epsilon^{\mu\nu\rho\sigma\mu_1\cdots \mu_6}}{\sqrt{-g}}
        H_{\mu\mu_1\mu_2} F_{\mu_3\cdots \mu_6},
\label{EQ4} \\
&& R_{\mu\nu} - \frac12 g_{\mu\nu} R =
      - \frac14 g_{\mu\nu} \nabla_\rho \Phi \nabla^\rho \Phi
      + \frac12 \nabla_\mu\Phi\nabla_\nu\Phi
\nonumber \\
&&\hskip 29mm
      - \frac1{24} g_{\mu\nu} e^{-\Phi}
             H_{\rho\sigma\lambda} H^{\rho\sigma\lambda}
      + \frac14 e^{-\Phi} H_{\mu\rho\sigma} H_\nu{}^{\rho\sigma}
\nonumber \\
&&\hskip 29mm
      -\frac12 g_{\mu\nu}
       \left(  \frac14 e^{\frac32\Phi} F_{\rho\sigma} F^{\rho\sigma}
             + \frac1{48} e^{\frac12\Phi}
                   \tilde{F}_{\rho\sigma\lambda\kappa}
                   \tilde{F}^{\rho\sigma\lambda\kappa}
       \right)
\nonumber \\
&&\hskip 29mm
     + \frac12 e^{\frac32 \Phi} F_{\mu\rho} F_\nu{}^\rho
     + \frac1{12} e^{\frac12 \Phi}
                \tilde{F}_{\mu\rho\sigma\lambda}
                \tilde{F}_\nu{}^{\rho\sigma\lambda},
\label{EQ5}
\end{eqnarray}
where $\epsilon^{\mu_1\cdots \mu_{10}}$ is the Levi-Civita symbol
chosen by $\epsilon^{+-12\cdots 8} = 1$, and the field strengths are
defined as
\begin{eqnarray}
F_{\mu\nu} &=& 2 \partial_{[\mu} A_{\nu]}, \nonumber \\
H_{\mu\nu\rho} &=& 3 \partial_{[\mu} B_{\nu\rho]},  \nonumber \\
F_{\mu\nu\rho\sigma} &=& 4 \partial_{[\mu} A_{\nu\rho\sigma]}, \nonumber \\
\tilde F_{\mu\nu\rho\sigma} &=& 4 \partial_{[\mu}
A_{\nu\rho\sigma]} + 4 A_{[\mu} H_{\nu\rho\sigma]}.  \nonumber
\end{eqnarray}

To obtain the linearized equations of motion for the fluctuation
fields, we expand the fields near the pp-wave background given
in Eq.~(\ref{pp-wave}) as follows
\begin{eqnarray}
&&\Phi = \phi, \nonumber \\
&&g_{\mu\nu} = \bar g_{\mu\nu} + h_{\mu\nu} \rightarrow R_{\mu\nu}
= \bar R_{\mu\nu}  + r_{\mu\nu},
\nonumber \\
&&A_\mu = \bar A_\mu + a_\mu \rightarrow F_{\mu\nu} = \bar
F_{\mu\nu} + f_{\mu\nu},
\nonumber \\
&& B_{\mu\nu} = b_{\mu\nu} \rightarrow H_{\mu\nu\rho} =
h_{\mu\nu\rho},
\nonumber \\
&& A_{\mu\nu\rho} = \bar A_{\mu\nu\rho} + a_{\mu\nu\rho}
\rightarrow
       F_{\mu\nu\rho\sigma} = \bar F_{\mu\nu\rho\sigma}
                             + f_{\mu\nu\rho\sigma},
\end{eqnarray}
where the fields with bar denote the background fields.  And we
shall choose the usual light-cone gauge for the fluctuations,
$a_\mu, b_{\mu\nu}, a_{\mu\nu\rho}, h_{\mu\nu}$ such as
\begin{equation}\label{gfix}
a_{-} =  b_{-I} =  a_{-IJ} =  h_{-I} = 0.
\end{equation}

The linearized form of the equation of motion (\ref{EQ1}) for
dilaton is the following coupled equation
\begin{equation} \label{LEQ1}
\Box \phi = \frac{\mu}2 \partial_- ( a_4 - a_{123}).
\end{equation}
And the equations of motion (\ref{EQ2}), (\ref{EQ3}) and (\ref{EQ4})
for the gauge fields have the following linearized forms respectively,
\begin{eqnarray}
&& \partial_\mu
   \bigg(  \frac32 \phi \bar{F}^{\mu\nu}
         + f^{\mu\nu} - h^{\mu\mu'} \bar{F}_{\mu'}{}^\nu
         - \bar{F}^\mu{}_{\nu'} h^{\nu'\nu}
   \bigg) =
  -\frac1{3!} h_{\mu\rho\sigma} \bar{F}^{\mu\nu\rho\sigma} ~,
\label{LEQ2} \\
&& \partial_\mu
   \bigg( h^{\mu\nu\rho}
        + a_\sigma\bar F^{\sigma\mu\nu\rho}
        + \bar{A}_\sigma f^{\sigma\mu\nu\rho}
        + 4 \bar{A}_\sigma \bar{A}^{[\sigma} h^{\mu\nu\rho]}
   \bigg) =
  \frac{\mu}{4!}
  \epsilon^{+123\nu\rho\mu_1\cdots \mu_4} f_{\mu_1\cdots \mu_4} ~,
\label{LEQ3} \\
&& \partial_\mu
   \bigg( \frac12 \phi \bar{F}^{\mu\nu\rho\sigma}
         + f^{\mu\nu\rho\sigma}
         + 4 \bar{A}^{[\mu} h^{\nu\rho\sigma]}
         - h^{\mu\mu'} \bar{F}_{\mu'}{}^{\nu\rho\sigma}
\nonumber \\
&&  \qquad
         + h^{\nu\nu'} \bar{F}_{\nu'}{}^{\mu\rho\sigma}
         - h^{\rho\rho'} \bar{F}_{\rho'}{}^{\mu\nu\sigma}
         + h^{\sigma\sigma'} \bar{F}_{\sigma'}{}^{\mu\nu\rho}
    \bigg) =
        \frac{\mu}{3!}
        \epsilon^{+123\nu\rho\sigma\mu_1\mu_2\mu_3}
        h_{\mu_1\mu_2\mu_3} ~,
\nonumber
\\ \label{LEQ4}
\end{eqnarray}
where the raising and lowering of the Lorentz indices are
performed by the plane-wave metric $\bar g_{\mu\nu}$.

The ($+$),($+I$),($+IJ$) components of the Eqs.~(\ref{LEQ2}),
(\ref{LEQ3}) and (\ref{LEQ4}) give constraints implying that the modes
$a_+$, $b_{+I}$ and $a_{+IJ}$ are non-dynamical.  We can express these
non-dynamical modes in terms of physical modes $a_I$, $b_{IJ}$,
$a_{IJK}$ as follows,
\begin{eqnarray}
a_+ &=& \frac1{\partial_-}\partial_I a_I,\label{const1} \\
b_{I+} &=& \frac1{\partial_-}\partial_J b_{IJ}, \label{const2} \\
a_{IJ+} &=& \frac1{\partial_-} \left( \partial_K a_{IJK}
            + \bar A_+\partial_- b_{IJ}\right), \label{const3}
\end{eqnarray}
where $\bar A_+ = \frac{\mu}3 x_4$.

The linearized form of Einstein equation (\ref{EQ5}) has rather
complicated form,
\begin{eqnarray}
r_{\mu\nu} - \frac12 \bar g_{\mu\nu} r
  &=& - \frac{\mu}6 \bar{g}_{\mu\nu} \partial_- a_4
      + \frac{\mu}2 \bar{g}_{\mu\nu} \partial_- a_{123}
\nonumber \\
  & & + \frac34 \phi \bar{F}_{\mu\rho} \bar{F}_\nu{}^\rho
      + \frac1{24} \phi \bar{F}_{\mu\rho\sigma\lambda}
          \bar{F}_\nu{}^{\rho\sigma\lambda}
      + \frac12 \left( \bar{F}_{\mu\rho} f_\nu{}^\rho
                      + f_{\mu\rho} \bar{F}_\nu{}^\rho
                \right)
\nonumber \\
  & & + \frac1{12}
        \left[ \bar{F}_\mu{}^{\rho\sigma\lambda}
               \left( f_{\nu\rho\sigma\lambda}
                     + 4 \bar{A}_{[\nu} h_{\rho\sigma\lambda]}
               \right)
             + \left( f_{\mu\rho\sigma\lambda}
                     + 4 \bar{A}_{[\mu} h_{\rho\sigma\lambda]}
               \right)
               \bar{F}_\nu{}^{\rho\sigma\lambda} \right]
\nonumber \\
  & & - \frac12 \bar{F}_{\mu\rho}
              \bar{F}_{\nu\rho'} h^{\rho\rho'}
      - \frac14 \bar{F}_{\mu\rho\sigma\lambda}
                \bar{F}_{\nu\rho'}{}^{\sigma\lambda}
                h^{\rho\rho'}
\label{LEQ5}
\end{eqnarray}
with
\begin{eqnarray}
r_{\mu\nu} &=&
      \frac12
      \left( -\bar \nabla^2 h_{\mu\nu}
           +\bar \nabla_\mu \bar\nabla^\rho h_{\rho\nu}
           + \bar\nabla_\nu \bar \nabla^\rho h_{\rho\mu}
           - \bar\nabla_\mu \bar\nabla_\nu h^\rho_{\rho}
      \right.
\nonumber \\
&& \hskip 5mm
  \left.
            + 2 \bar R_{\mu\rho\sigma\nu} h^{\rho\sigma}
            + \bar R_{\mu\rho} h^\rho_\nu
            + \bar R_{\nu\rho} h^\rho_\mu ~
  \right) ~,
\label{flucR} \\
r &\equiv& \bar g^{\mu\nu} r_{\mu\nu}, \label{flucRR}
\end{eqnarray}
where the covariant derivative $\bar\nabla_\mu$ is defined by using
the background pp-wave metric $\bar g_{\mu\nu}$ and the non-vanishing
connection and curvature quantities are given by
\begin{equation}
\Gamma^-_{I+} = \frac{1}{2} \partial_I A,\quad
\Gamma^I_{++} = \frac{1}{2} \partial_I A,\quad
R_{+IJ+} = \frac{1}{2} \partial_I \partial_J A, \quad
R_{++} = \frac{1}{2} \partial^2_I A ~.
\label{CC}
\end{equation}
The ($--$) component of the Eq.~(\ref{LEQ5}) gives the following
zero-trace condition for the transverse modes of the graviton
\begin{equation}\label{const4}
h_{II} = 0.
\end{equation}
And the ($-I$) components of the Eq.~(\ref{LEQ5}) lead to the
expressions for non-dynamical modes $h_{+I}$,
\begin{equation}\label{const5}
 h_{+I} = \frac1{\partial_-}  \partial_J h_{IJ}.
\end{equation}
Under the conditions (\ref{const4}) and (\ref{const5}), the trace for
the space-time indices of the Eq.~(\ref{LEQ5}) gives an additional
condition,
\begin{equation}\label{const6}
h_{++} = \frac1{(\partial_-)^2} \partial_I  \partial_J h_{IJ}
 + \frac{\mu}4  \frac1{\partial_-}(a_4 - a_{123}).
\end{equation}

Now let us consider the linearized equations for physical modes which
determine the light-cone energy spectrum.  There are four sets of
decoupled modes and seven sets of coupled ones which we need to
diagonalize to determine the light-cone energy spectrum for the
physical modes.  The linearized equation for dilaton, (4)-component of
the Eq.~(\ref{LEQ2}), (123)-component of the Eq.~(\ref{LEQ4}), the
trace for $SO(3)$ indices and (44)-component of the Eq.~(\ref{LEQ5})
are coupled each other and given by\footnote{From now one, the index
  of the type $i$ is taken to run from $1$ to $3$, while the range of
  $i'$ is not changed.}
\begin{eqnarray}
&& \Box \phi - \frac{\mu}2\partial_- a_4
             + \frac{\mu}2 \partial_- a_{123} = 0 ~,~~~
   \Box a_4 + \frac{\mu}2\partial_-\phi
             - \frac{\mu}3 \partial_- h_{44} = 0,
\nonumber \\
&& \Box a_{123} - \frac{\mu}2\partial_- \phi
               + \mu \partial_- h_{ii} = 0 ~,~~~
   \Box h_{ii} - \frac{\mu}4 \partial_- a_4
               - \frac{15\mu}4 \partial_- a_{123} = 0 ~,
\nonumber \\
&& \Box h_{44} + \frac{7\mu}{12} \partial_- a_4
               + \frac{3\mu}4\partial_- a_{123} = 0.
\label{coup1}
\end{eqnarray}
These coupled equations form $SO(3)\times SO(4)$ scalar multiplet.
Then we obtain the following diagonalized equations
\begin{eqnarray}
& & \Box \phi_0 = 0 ~,
  \nonumber \\
& & \Box \phi_2 + i \frac{2\mu}{3}
        \partial_- \phi_2 = 0 ~,~~~
    \Box \bar\phi_2 - i \frac{2\mu}{3}
        \partial_- \bar\phi_2 = 0 ~,
  \nonumber \\
& & \Box\phi_6 + 2i \mu \partial_- \phi_6 = 0 ~,~~~
    \Box \bar\phi_6 - 2i \mu \partial_- \bar\phi_6 = 0,
\label{dia1}
\end{eqnarray}
where we define
\begin{eqnarray}
\phi_0 &\equiv&  \phi + \frac13 h_{ii} + h_{44} ~,
  \nonumber \\
\phi_2 &\equiv& \phi + \frac43 i a_4 - \frac23 h_{44} ~,~~~
\bar\phi_2 \equiv \phi - \frac43 i a_4 - \frac23 h_{44} ~,
  \nonumber \\
\phi_6 &\equiv& \phi - 4 i a_{123} - 2 h_{ii} ~,~~~
    \bar\phi_6 \equiv \phi + 4 i a_{123} - 2 h_{ii} ~.
\label{mod1}
\end{eqnarray}
According to Eqs.~(\ref{seqn}) and (\ref{lstE}), these equations in
Eq.~(\ref{dia1}) mean that the minimal light-cone energies of the
fields in Eq.~(\ref{mod1}) are given by
\begin{equation}\label{eng1}
{\cal E}_0\left( \phi_0 \right) = 6 ~,~~~
{\cal E}_0\left(\phi_2 \right) = 4 ~,~~~
{\cal E}_0\left( \bar\phi_2\right) = 8 ~,~~~
{\cal E}_0\left( \phi_6 \right) = 0 ~,~~~
{\cal E}_0\left( \bar\phi_6 \right) = 12 ~.
\end{equation}

There are two sets of $SO(3)$ vector multiplets. One is decoupled
multiplet and the other is coupled one. The decoupled one comes from
the ($i4$)-components of the Eq.~(\ref{LEQ3}) and given by
\begin{equation}\label{decoup1}
\Box \beta_{0i} = 0,
\end{equation}
where $\beta_{0i} \equiv b_{4i}$ and their
minimal light-cone energies are
\begin{equation}
{\cal E}_0\left(\beta_{0i}\right) = 6~.
\end{equation}
The coupled $SO(3)$ vector multiplets come from the ($i$), ($ij$),
($4ij$) and ($i4$)-components of the Eqs.~(\ref{LEQ2}),
(\ref{LEQ3}), (\ref{LEQ4}) and (\ref{LEQ5}) respectively and
written by
\begin{eqnarray}
&& \Box a_i - \frac{\mu}3 \partial_- h_{4i}
         - \frac{\mu}2 \epsilon_{ijk}\partial_- b_{jk} = 0 ~,~~~
   \Box b_{ij} + \mu \epsilon_{ijk}\partial_- a_k
         - \frac{\mu}3 \partial_- a_{4ij} = 0,
 \nonumber  \\
&& \Box a_{4ij} + \frac{\mu}3 \partial_- b_{ij}
         + \mu \epsilon_{ijk} \partial_- h_{4k} = 0 ~,~~~
   \Box h_{4i}  + \frac{\mu}3 \partial_- a_i
         - \frac{\mu}2 \epsilon_{ijk}\partial_- a_{4jk} = 0 ~.
\label{coup2}
\end{eqnarray}
These coupled equations are diagonalized as
\begin{eqnarray}
&& \Box \beta_{2i}
        + i \frac{2\mu}{3} \partial_- \beta_{2i} = 0 ~,~~~
   \Box \bar\beta_{2i}
        - i \frac{2\mu}{3} \partial_- \bar\beta_{2i} = 0 ~,
 \nonumber  \\
&& \Box \beta_{4i}
        + i \frac{4\mu}{3} \partial_- \beta_{4i} = 0 ~,~~~
   \Box \bar\beta_{4i}
        - i \frac{4\mu}{3} \partial_- \bar\beta_{4i} = 0 ~,
\label{dia2}
\end{eqnarray}
where we define the diagonalized physical modes as
\begin{eqnarray}
\beta_{2i} &\equiv& a_i - i h_{4i} + \frac{i}2\epsilon_{ijk}
b_{jk}
               + \frac12\epsilon_{ijk}a_{4jk},
\quad \bar\beta_{2i} \equiv a_i + i h_{4i} -
\frac{i}2\epsilon_{ijk} b_{jk}
                + \frac12\epsilon_{ijk}a_{4jk},
\nonumber \\
\beta_{4i} &\equiv& a_i + i h_{4i} + \frac{i}2\epsilon_{ijk}
b_{jk}
                - \frac12\epsilon_{ijk}a_{4jk},
\quad \bar\beta_{4i} \equiv a_i - i h_{4i} -
\frac{i}2\epsilon_{ijk} b_{jk}
               - \frac12\epsilon_{ijk}a_{4jk},
\label{mod2}
\end{eqnarray}
whose minimal light-cone energies are given by
\begin{equation}\label{eng2}
{\cal E}_0\left( \beta_{2i}\right) = 4 ~,~~~
{\cal E}_0\left( \bar\beta_{2i}\right) = 8 ~,~~~
{\cal E}_0\left( \beta_{4i}\right) = 2 ~,~~~
{\cal E}_0\left( \bar\beta_{4i}\right) = 10~.
\end{equation}

There are two kinds of coupled multiplets in $SO(4)$ vector
multiplets. The ($i'$) and ($4i'$)-components of Eqs.~(\ref{LEQ2}) and
(\ref{LEQ3}) form one set of coupled equations
\begin{eqnarray}
\Box a_{i'} - \frac{\mu}3 \partial_- h_{4i'} = 0, \qquad   \Box
h_{4i'} + \frac{\mu}3 \partial_- a_{i'} = 0, \label{coup3}
\end{eqnarray}
and the ($4i'$) and ($i'j'k'$)-components of Eqs.~(\ref{LEQ3}) and
(\ref{LEQ4}) give the other set
\begin{eqnarray}
 \Box b_{4i'} + \frac{\mu}6 \epsilon_{i'j'k'l'}
                \partial_- a_{j'k'l'} = 0,
\qquad \Box a_{i'j'k'} + \mu \epsilon_{i'j'k'l'}
                  \partial_- b_{4l'} = 0,
\label{coup4}
\end{eqnarray}
where $\epsilon_{i'j'k'l'}$ are Levi-Civita symbols and we choose
$\epsilon_{5678} = 1$.  By introducing two sets of complex $SO(4)$
vectors
\begin{eqnarray}
&&\beta_{1i'} \equiv a_{i'} +  i h_{4i'},\qquad \bar\beta_{1i'}
\equiv a_{i'} - i h_{4i'},
\label{mod3} \\
&& \beta_{3i'} \equiv b_{4i'}
                    - \frac{i}6 \epsilon_{i'j'k'l'} a_{j'k'l'},
\qquad \bar \beta_{3i'} \equiv b_{4i'}
                    + \frac{i}6 \epsilon_{i'j'k'l'} a_{j'k'l'},
\label{mod4}
\end{eqnarray}
we can diagonalize the Eqs.~(\ref{coup3}) and (\ref{coup4}) as
\begin{eqnarray}
&&\Box \beta_{1i'}
     + i \frac{\mu}{3} \partial_- \beta_{1i'} = 0 ~,~~~
  \Box \bar\beta_{1i'}
     - i \frac{\mu}{3} \partial_- \bar\beta_{1i'} = 0 ~,
\label{dia3} \\
&&\Box \beta_{3i'} + i \mu \partial_- \beta_{3i'} = 0 ~,~~~
  \Box \bar \beta_{3i'} - i \mu \partial_- \bar\beta_{3i'} = 0 ~.
\label{dia4}
\end{eqnarray}
Thus we obtain the minimal light-cone energies of the diagonalized
modes as
\begin{equation}\label{eng3}
{\cal E}_0 (\beta_{1i'}) = 5 ~,~~~
{\cal E}_0 (\bar\beta_{1i'}) = 7 ~,~~~
{\cal E}_0 (\beta_{3i'}) = 3 ~,~~~
{\cal E}_0 (\bar\beta_{3i'}) = 9 ~.
\end{equation}

The ($ij'$)-components of the Eq.~(\ref{LEQ3}) and the
($4ij'$)-components of the Eq.~(\ref{LEQ4}) form a coupled set of
equations as follows,
\begin{equation}\label{coup5}
\Box b_{ij'} - \frac{\mu}3 \partial_- a_{4ij'} = 0, \qquad \Box
a_{4ij'} + \frac{\mu}3\partial_- b_{ij'} = 0.
\end{equation}
By defining the complex tensors
\begin{equation}\label{mod5}
\beta_{1ij'} \equiv b_{ij'} + i a_{4ij'}, \qquad \bar\beta_{1ij'}
\equiv b_{ij'} - i a_{4ij'},
\end{equation}
the equations in Eq.~(\ref{coup5}) are diagonalized as
\begin{equation}\label{dia5}
\Box \beta_{1ij'}
   + i \frac{\mu}{3} \partial_- \beta_{1ij'} = 0 ~,~~~
\Box \bar\beta_{1ij'}
   - i \frac{\mu}{3} \partial_- \bar\beta_{1ij'} = 0 ~,
\end{equation}
so that the minimal light-cone energies are
\begin{equation}\label{eng5}
{\cal E}_0(\beta_{1ij'}) = 5 ~,~~~
{\cal E}_0(\bar\beta_{1ij'}) = 7 ~.
\end{equation}

The ($i'j'$)-components of the Eq.~(\ref{LEQ3}) and ($4i'j'$)-ones of
the Eq.~(\ref{LEQ4}) are coupled also and form anti-symmetric 2-form
field multiplets. The coupled equations are
\begin{eqnarray}
&& \Box b_{i'j'} - \frac{\mu}3 \partial_- a_{4i'j'}
                + \frac{\mu}2 \epsilon_{i'j'k'l'}
                  \partial_- a_{4k'l'} = 0,
\nonumber \\
&&\Box a_{4i'j'} + \frac{\mu}3 \partial_- b_{i'j'}
                 - \frac{\mu}2 \epsilon_{i'j'k'l'}
                   \partial_- b_{k'l'} = 0,
\label{coup6}
\end{eqnarray}
and these are diagonalized by defining the anti-symmetric complex
2-form fields,
\begin{eqnarray}
\beta_{2i'j'} &\equiv& a^+_{i'j'} - i {\tilde a^+}_{i'j'}, \qquad
\bar\beta_{2i'j'} \equiv a^+_{i'j'} + i {\tilde a^+}_{i'j'},
\nonumber \\
\beta_{4i'j'} &\equiv& a^-_{i'j'} + i {\tilde a^-}_{i'j'}, \qquad
\bar\beta_{4i'j'} \equiv a^-_{i'j'} - i {\tilde a^-}_{i'j'},
\label{mod6}
\end{eqnarray}
where we used the (anti-)self-dual tensors which are irreducible
tensors of the $SO(4)$ algebra and defined by
\begin{eqnarray}
a^{\pm}_{i'j'} \equiv b_{i'j'}
                           \pm \frac12\epsilon_{i'j'k'l'}
                            b_{k'l'},
\qquad {\tilde a^{\pm}}_{i'j'} \equiv  a_{4i'j'}
                                \pm \frac12\epsilon_{i'j'k'l'}
                                a_{4k'l'}.
\nonumber
\end{eqnarray}
Then the equations in Eq.~(\ref{coup6}) are diagonalized by
\begin{eqnarray}
&&\Box \beta_{2i'j'}
    + i \frac{2\mu}{3} \partial_- \beta_{2i'j'} = 0 ~,~~~
  \Box \bar\beta_{2i'j'}
    - i \frac{2\mu}{3} \partial_- \bar\beta_{2i'j'} = 0 ~,
\nonumber  \\
&&\Box \beta_{4i'j'}
    + i \frac{4\mu}{3} \partial_- \beta_{4i'j'} = 0 ~,~~~
  \Box \bar\beta_{4i'j'}
    - i \frac{4\mu}{3} \partial_- \bar\beta_{4i'j'} = 0 ~,
\label{dia6}
\end{eqnarray}
thus the minimal light-cone energies are
\begin{equation}\label{eng6}
{\cal E}_0(\beta_{2i'j'}) = 4 ~,~~~
{\cal E}_0(\bar\beta_{2i'j'}) = 8 ~,~~~
{\cal E}_0(\beta_{4i'j'}) = 2 ~,~~~
{\cal E}_0(\bar\beta_{4i'j'}) = 10 ~.
\end{equation}

There is remaining one mixed multiplet. The ($ijk'$)-components of the
Eq.~(\ref{LEQ4}) and ($ij'$)-ones of the Eq.~(\ref{LEQ5}) give
\begin{eqnarray}
 \Box a_{ijk'} + \mu \epsilon_{ijl}
                  \partial_- h_{lk'} = 0,
\qquad
 \Box h_{ij'} - \frac{\mu}2 \epsilon_{ikl}
\partial_- a_{klj'} = 0.
\label{coup7}
\end{eqnarray}
By defining the complex tensors
\begin{eqnarray}
\beta_{3ijk'} \equiv a_{ijk'} - i \epsilon_{ijk} h_{kk'}, \qquad
\bar\beta_{3ijk'} \equiv a_{ijk'} + i \epsilon_{ijk} h_{kk'}, \label{mod7}
\end{eqnarray}
we can diagonalize the equations in Eq.~(\ref{coup7}) as
\begin{eqnarray}
\Box \beta_{3ijk'} + i\mu\partial_- \beta_{3ijk'} = 0, \qquad \Box
\bar\beta_{3ijk'} - i\mu\partial_- \bar\beta_{3ijk'} = 0, \label{dia7}
\end{eqnarray}
so that the minimal energies are
\begin{equation}\label{eng7}
{\cal E}_0(\beta_{3ijk'}) = 3 ~,~~~
{\cal E}_0(\bar\beta_{2ijk'}) = 9 ~.
\end{equation}

The ($ij'k'$)-components of the Eq.~(\ref{LEQ4}) are decoupled and
the linearized equations and minimal energies are given by
\begin{equation}\label{decoup2}
\Box \beta_{0ij'k'} = 0 ~,~~~
{\cal E}_0 (\beta_{0ij'k'}) = 6 ~,
\end{equation}
where $\beta_{0ij'k'} \equiv a_{ij'k'}$.

From ($ij$) and ($i'j'$)-components of the Eq.~(\ref{LEQ5}),
we can extract two sets of traceless gravitons which belong to
$SO(3)$ and $SO(4)$ graviton multiplets respectively. Then the
equations and minimal light-cone energies are given by
\begin{equation}\label{decoup3}
\Box h^{\perp}_{ij} = 0, \qquad \Box h^{\perp}_{i'j'} =0 ~,~~~
{\cal E}_0 (h^{\perp}_{ij})
          = {\cal E}_0(h^{\perp}_{i'j'}) = 6 ~,
\end{equation}
where we have defined
\begin{eqnarray}
h^{\perp}_{ij} \equiv h_{ij} - \frac13 \delta_{ij} h_{kk}, \qquad
h^{\perp}_{i'j'} \equiv h_{i'j'} - \frac14 \delta_{i'j'}h_{k'k'}.
\nonumber
\end{eqnarray}

\subsection{Fermionic excitations}
\label{f-sugra}

The fermionic fields of Type IIA supergravity are spin-1/2 dilatino
$\Lambda$ and spin-3/2 gravitino $\Psi_\mu$, each of which has real 32
components and is decomposed into two pieces of opposite $SO(1,9)$
chiralities.  Though it is usual to decompose the fermionic fields
based on $SO(1,9)$ chiralities, it will be convenient to take a
different decomposition in this paper in a way that the $SO(3) \times
SO(4)$ symmetry structure of the pp-wave background manifests.

If we do not take any decomposition for a while, the equations of
motion for $\Lambda$ and $\Psi_\mu$ are as follows:
\begin{eqnarray}
\Gamma^\mu D_\mu \Lambda
 &=& - \frac{\sqrt{2}}{4} \Gamma^9 \Gamma^\mu \Gamma^\nu
        \Psi_\mu \partial_\nu \Phi
     - \frac{1}{192} e^{\Phi/4}
      \left( \frac{1}{\sqrt{2}} \Gamma^9 \Gamma^\mu
          \Gamma^{\rho\sigma\lambda\kappa} \Psi_\mu
       - \frac{3}{2} \Gamma^{\rho\sigma\lambda\kappa} \Lambda
      \right)
      {\tilde{F}}_{\rho\sigma\lambda\kappa}
  \nonumber \\
 & & + \frac{\sqrt{2}}{48} e^{-\Phi/2} \Gamma^\mu
       \Gamma^{\rho\sigma\lambda} \Psi_\mu H_{\rho\sigma\lambda}
     - \frac{1}{16} e^{3 \Phi /4}
     \left( \frac{3}{\sqrt{2}} \Gamma^\mu \Gamma^{\rho\sigma} \Psi_\mu
      + \frac{5}{2} \Gamma^9 \Gamma^{\rho\sigma} \Lambda
     \right)  F_{\rho\sigma}
  \nonumber \\
 & & + \cdots ~,
\label{dilatino}
\end{eqnarray}
\begin{eqnarray}
\Gamma^{\mu\nu\rho} D_\nu \Psi_\rho
 &=&  \frac{\sqrt{2}}{4} \Gamma^\nu \Gamma^\mu \Gamma^9
            \Lambda \partial_\nu \Phi
  \nonumber \\
 & & - \frac{1}{192} e^{\Phi/4}
      \bigg( 2 \Gamma^{\mu\nu\rho\sigma\lambda\kappa} \Psi_\nu
           + 24 g^{\mu\rho} \Gamma^{\sigma\lambda} \Psi^\kappa
           + \frac{1}{\sqrt{2}} \Gamma^{\rho\sigma\lambda\kappa}
             \Gamma^\mu \Gamma^9 \Lambda
      \bigg)  {\tilde{F}}_{\rho\sigma\lambda\kappa}
  \nonumber \\
 & & + \frac{1}{48} e^{-\Phi/2}
      \bigg( 2 \Gamma^9 \Gamma^{\mu\nu\rho\sigma\lambda} \Psi_\nu
            -12 g^{\mu\rho} \Gamma^9 \Gamma^\sigma \Psi^\lambda
            + \sqrt{2} \Gamma^{\rho\sigma\lambda} \Gamma^\mu
              \Lambda
      \bigg) H_{\rho\sigma\lambda}
  \nonumber \\
 & & + \frac{1}{16} e^{3 \Phi /4}
     \bigg( 2 \Gamma^9 \Gamma^{\mu\nu\rho\sigma} \Psi_\nu
           + 4 g^{\mu\rho} \Gamma^9 \Psi^\sigma
           - \frac{3}{\sqrt{2}} \Gamma^{\rho\sigma} \Gamma^\mu
             \Lambda
     \bigg) F_{\rho\sigma} + \cdots
  \nonumber \\
 & \equiv & J^\mu + \cdots ~,
\label{gravitino}
\end{eqnarray}
where the dots on the right hand sides represent the terms of cubic in
fermionic fields, which describe the interactions between excitations
and hence are not our concern, and the covariant derivative for spinor
is given by\footnote{The indices $r,s,...$ represent the flat tangent
  space indices.  The gamma matrices $\Gamma^r$ satisfy the $SO(1,9)$
  Clifford algebra, $\{ \Gamma^r, \Gamma^s \} = 2 \eta^{rs}$, where
  $\eta^{rs}$ is the ten dimensional flat metric. We note that all the
  indices of gamma matrices, which are not denoted by Greek
  characters, are flat indices.}
\begin{equation}
D_\mu = \partial_\mu + \frac{1}{4} \omega_\mu^{rs} \Gamma_{rs} ~,
\end{equation}
whose explicit expression for the pp-wave background, (\ref{pp-wave}),
is
\begin{equation}
D_+ = \partial_+ - \frac{1}{4} \partial_I A \Gamma^{+I} ~,~~~
D_- = \partial_- ~,~~~
D_I = \partial_I ~,
\end{equation}
under the following choice of the zehnbein
\begin{equation}
e^+ = dx^+ ~,~~~
e^- = dx^- + \frac{1}{2} A(x^I) dx^+ ~,~~~
e^I = dx^I ~.
\label{pp}
\end{equation}
For the study of the linearized equation of motion for gravitino, we
note that it is convenient to rewrite the Eq.~(\ref{gravitino}) as
\begin{equation}
\Gamma^\nu  D_\nu \Psi_\mu - D_\mu \Psi
 = J_\mu - \frac{1}{8} \Gamma_\mu \Gamma_\nu J^\nu ~,
\label{s3/2}
\end{equation}
where $\Psi \equiv \Gamma^\mu \Psi_\mu$ and we have ignored the
interaction terms represented by dots in Eq.~(\ref{gravitino}).  For
the consideration of the physical modes, we should impose the gauge
condition for the gravitino, which we take as the following light-cone
gauge,
\begin{equation}
\Psi_- = 0 ~.
\label{f-gauge}
\end{equation}

In the light-cone formulation, the fermionic fields are decomposed into
the dynamical (physical) and non-dynamical modes explicitly by using
$\{ \Gamma^+ , \Gamma^- \} = 2 \eta^{+-}$.  For the spin-1/2 field, we
have the decomposition,
\begin{equation}
\Lambda = \lambda + \eta ~,
\label{d1/2}
\end{equation}
where
\begin{equation}
\lambda \equiv - \frac{1}{2} \Gamma^- \Gamma^+ \Lambda ~,~~~
\eta \equiv - \frac{1}{2} \Gamma^+ \Gamma^- \Lambda ~,
\label{d1/2-d}
\end{equation}
while for the spin-3/2 gravitino field,
\begin{equation}
\Psi_\mu = \psi_\mu + \rho_\mu~,
\label{d3/2}
\end{equation}
where
\begin{equation}
\psi_\mu \equiv - \frac{1}{2} \Gamma^- \Gamma^+ \Psi_\mu ~,~~~
\rho_\mu \equiv - \frac{1}{2} \Gamma^+ \Gamma^- \Psi_\mu ~.
\label{d3/2-d}
\end{equation}
As we will see, $\eta$ and $\rho_\mu$ correspond to non-dynamical
fields expressed in terms of the physical fields $\lambda$ and
$\psi_\mu$ respectively.

Having the decomposition of the fermionic fields, Eqs.~(\ref{d1/2})
and (\ref{d3/2}), we first consider the equation of motion for the
spin-1/2 field $\Lambda$, Eq.~(\ref{dilatino}).  The linearized
form of the equation of motion is
\begin{eqnarray}
\lefteqn{ \Gamma^+ ( \partial_+ - \frac{1}{2} A \partial_- ) \lambda
       + \Gamma^- \partial_- \eta
       + \Gamma^I \partial_I (\lambda + \eta) }
    \nonumber \\
 &=& \frac{\mu}{48} \Gamma^+ \Gamma^{49} (5 - 9 \Gamma^{12349} ) \lambda
     - \frac{\mu}{8 \sqrt{2}} \Gamma^+ \Gamma^I \Gamma^4
       ( 1 - \Gamma^{12349} ) \psi_I ~.
\end{eqnarray}
We see that the field $\eta$ does not have the dependence on the
light-cone time $x^+$ and is expected to be non-dynamical.  Indeed,
this equation leads to the expression of $\eta$ in terms of $\lambda$
as
\begin{equation}
\eta = \frac{1}{2 \partial_-} \Gamma^+ \Gamma^I \partial_I \lambda ~.
\end{equation}
Taking into account this expression, the linearized equation of motion
for the physical mode $\lambda$ becomes
\begin{equation}
\Box \lambda =
- \frac{\mu}{24} \Gamma^{49} ( 5 - 9 \Gamma^{12349} )
       \partial_- \lambda
+ \frac{\mu}{4 \sqrt{2}} \Gamma^I \Gamma^4
     ( 1 - \Gamma^{12349} ) \partial_- \psi_I ~.
\label{dil-eq}
\end{equation}

We see that the gravitino appears in the equation of motion for the
dilatino, (\ref{dil-eq}).  This lets us turn to the equation of motion
for the gravitino and pick up the physical modes before going on
further analysis for the dilatino.  The explicit expressions for the
`current' $J_\mu$ are first needed in the light-cone gauge,
(\ref{f-gauge}).  Except for the light-cone component, $J_+$, we have
$J_- = 0$ and, for the spatial components,
\begin{eqnarray}
J_i &=& - \frac{\mu}{4} \Gamma^+
   \bigg( \Gamma^{123} ( \delta^{ij}-\Gamma^i \Gamma^j ) \psi_j
         - \frac{1}{3} \Gamma^{49} \psi_i
         + \frac{1}{3} \Gamma^{i9} \psi_4
         + \frac{1}{2\sqrt{2}} \Gamma^i \Gamma^4
           ( 1 - \Gamma^{12349} ) \lambda
   \bigg) ~,
  \nonumber \\
J_4 &=& - \frac{\mu}{4} \Gamma^+
    \bigg( \Gamma^{i'} \Gamma^{1234} \psi_{i'}
          - \frac{1}{2 \sqrt{2}} ( 1 - \Gamma^{12349} ) \lambda
    \bigg) ~,
  \nonumber \\
J_{i'} &=& \frac{\mu}{4} \Gamma^+
    \bigg(  \Gamma^{123} ( \delta^{i'j'}-\Gamma^{i'}\Gamma^{j'} )
               \psi_{j'}
           + \frac{1}{3} \Gamma^{49} \psi_{i'}
           + \Gamma^{i'} \Big( \Gamma^{1234} - \frac{1}{3} \Gamma^9
                       \Big)  \psi_4
  \nonumber \\
   & &     - \frac{1}{2\sqrt{2}} \Gamma^{i'} \Gamma^4
              ( 1 + \Gamma^{12349} ) \lambda
    \bigg) ~.
\label{j-current}
\end{eqnarray}

Then the $\mu = -$ component of Eq.~(\ref{s3/2}) gives
\begin{equation}
\Psi = \Gamma^+ \Psi_+ + \Gamma^I \Psi_I = 0 ~,
\label{m-}
\end{equation}
which implies $\Gamma^+ \Gamma^I \Psi_I = 0$ and states that $\Psi_+$
is non-dynamical.  For the $\mu = I$ component, we have
\begin{equation}
\Gamma^+ ( \partial_+ - \frac{1}{2} A \partial_- ) \psi_I
       + \Gamma^- \partial_- \rho_I
       + \Gamma^J \partial_J (\psi_I + \rho_I)
 = J_I - \frac{1}{8} \Gamma_I ( \Gamma^J J_J + \Gamma^+ J_+) ~,
\label{grav-s}
\end{equation}
with the decomposition, (\ref{d3/2}), and the expressions of $J_I$,
(\ref{j-current}).  We would like to note that the term $\Gamma^+ J_+$
on the right hand side of Eq.~(\ref{grav-s}) vanishes due to
Eq.~(\ref{m-}), and $J_+$ does not play any role in the following
formulation.  From (\ref{grav-s}), we now see that the $\rho_I$ field
is obviously non-dynamical, and is given in terms of $\psi_I$ as
\begin{equation}
\rho_I = \frac{1}{2 \partial_-}
        \Gamma^+ \Gamma^J \partial_J \psi_I ~.
\end{equation}
This relation leads us to the following linearized equation of motion
for the physical modes $\psi_I$:
\begin{equation}
\Box \psi_I - \Gamma^- \partial_-
(J_I - \frac{1}{8} \Gamma_I \Gamma^J J_J ) = 0 ~.
\label{gravi-eq}
\end{equation}
The remaining vector component of the gravitino is $\Psi_+$, which is
non-dynamical.  From Eqs.~(\ref{s3/2}) and (\ref{m-}), we have
\begin{equation}
\psi_+ = \frac{1}{2 \partial_-}
   ( \delta^{IJ} + \Gamma^{IJ} ) \partial_J \psi_I ~,~~~
\rho_+ = \frac{1}{2 \partial_-}
  ( \Gamma^+ \Gamma^I \partial_I \psi_+
   + \frac{1}{4} \Gamma^I J_I ) ~.
\end{equation}

As alluded at the beginning of this subsection, we now decompose the
physical modes to reflect the symmetry structure $SO(3) \times SO(4)$
of the pp-wave background.  However, to avoid some complexity, we are
concerned only about the $SO(4)$ structure.  The case of $SO(3)$ will
be required at later stage.  Since the decomposition is performed in
the transverse eight dimensional space and the physical mode has 16
independent spinor components, it is natural to use the $16 \times 16$
gamma matrices, $\gamma^I$, of Eq.~(\ref{gamma}).  In addition to
this, from our convention and the Eqs.~(\ref{d1/2-d}) and
(\ref{d3/2-d}), we make replacements $\lambda \rightarrow \left(
  \begin{array}{c} \lambda \\ 0
\end{array} \right)$ and $\psi_I \rightarrow \left( \begin{array}{c}
\psi_I \\ 0 \end{array} \right)$ to specify that the physical fields
have 16 components.  The decomposition of physical fields is then
\begin{equation}
\lambda = \lambda^+ + \lambda^- ~,~~~
\psi_I = \psi_I^+ + \psi_I^-
\end{equation}
where the sign of superscript indicates the $SO(4)$ chirality in the
space spanned by $x^5,...,x^8$, that is, $\gamma^{5678} \lambda^\pm =
\pm \lambda^\pm$ and $\gamma^{5678} \psi_I^\pm = \pm \psi_I^\pm$.  We
note that $\gamma^{5678} = \gamma^{12349}$, which will be useful in
the following formulation.

For each $SO(4)$ chirality, the dilatino equation of motion,
(\ref{dil-eq}), leads us to have
\begin{eqnarray}
 & & ( \Box - \frac{7 \mu}{12} \gamma^{49} \partial_- ) \lambda^+
    - \frac{\mu}{2\sqrt{2}} \gamma^{i4} \partial_- \psi^+_i
    - \frac{\mu}{2 \sqrt{2}} \partial_- \psi^+_4 = 0 ~,
\label{sd+}
     \\
 & & ( \Box + \frac{\mu}{6} \gamma^{49} \partial_- ) \lambda^-
    - \frac{\mu}{2 \sqrt{2}} \gamma^{i'4} \partial_- \psi^+_{i'} = 0
 ~.
\label{sd-}
\end{eqnarray}
For the case of gravitino, we first split the vector components as
$\psi_I^\pm = ( \psi_i^\pm, \psi_4^\pm, \psi_{i'}^\pm)$, and decompose
the modes $\psi_i^\pm$ and $\psi_{i'}^\pm$ into the transverse and
parallel parts with respect to the $\gamma^I$ matrices as follows:
\begin{eqnarray}
\psi^\pm_{\perp i} \equiv
( \delta_{ij} - \frac{1}{3} \gamma^i \gamma^j )
 \psi^\pm_j ~, & &
\psi^\pm_\| \equiv \gamma^i \psi^\pm_i ~,
   \nonumber \\
\psi^\pm_{\perp i'} \equiv ( \delta_{i'j'}
  - \frac{1}{4} \gamma^{i'} \gamma^{j'} )
 \psi^\pm_{j'} ~, & &
\psi'^\pm_\| \equiv \gamma^{i'} \psi^\mp_{i'} ~.
\end{eqnarray}
From Eq.~(\ref{gravi-eq}), we see that the equations of motion for the
transverse parts do not include other modes and are given by
\begin{eqnarray}
& & ( \Box - \frac{\mu}{3} \gamma^{123} \partial_- )
      \psi^+_{\perp i} = 0~,~~~
    ( \Box - \frac{2 \mu}{3}  \gamma^{123} \partial_-)
      \psi^-_{\perp i} = 0 ~,
 \nonumber \\
& & ( \Box + \frac{2 \mu}{3} \gamma^{123} \partial_- )
      \psi^+_{\perp i'} = 0~,~~~
    ( \Box + \frac{\mu}{3}  \gamma^{123} \partial_-)
      \psi^-_{\perp i'} = 0 ~.
\label{trans}
\end{eqnarray}
$\gamma^{123}$ in these equations now requires the modes to have
definite $SO(3)$ structure.  Since $(\gamma^{123})^2 = -1$, its
eigenvalues are $\pm i$, the $SO(3)$ chirality.  For a generic spinor
$\Theta$, we introduce the following notation.
\begin{equation}
\Theta^{\pm \pm} ~,
\label{34d}
\end{equation}
where the first superscript represents the SO(3) chirality and the
second the $SO(4)$ chirality.  Then the equations for the transverse
modes, (\ref{trans}), become
\begin{eqnarray}
& & ( \Box - i \frac{\mu}{3} \partial_- )
               \psi^{++}_{\perp i} = 0~,~~~
    ( \Box + i \frac{\mu}{3} \partial_- )
               \psi^{-+}_{\perp i} = 0~,~~~
 \nonumber \\
& & ( \Box - i \frac{2 \mu}{3} \partial_-)
               \psi^{+-}_{\perp i} = 0 ~,~~~
    ( \Box + i \frac{2 \mu}{3} \partial_-)
               \psi^{--}_{\perp i} = 0 ~,
 \nonumber \\
& & ( \Box + i \frac{2 \mu}{3} \partial_- )
               \psi^{++}_{\perp i'} = 0~,~~~
    ( \Box - i \frac{2 \mu}{3} \partial_- )
               \psi^{-+}_{\perp i'} = 0~,~~~
 \nonumber \\
& & ( \Box + i \frac{\mu}{3} \partial_-)
              \psi^{+-}_{\perp i'} = 0 ~,~~~
    ( \Box - i \frac{\mu}{3} \partial_-)
              \psi^{--}_{\perp i'} = 0 ~,
\end{eqnarray}
which state that, according to (\ref{lstE}), the minimal light-cone
energy for the respective physical gravitino modes are
\begin{equation}
{\cal E}_0 ( \psi^{--}_{\perp i}, \psi^{++}_{\perp i'} ) = 4 ~,~~~
{\cal E}_0 ( \psi^{-+}_{\perp i}, \psi^{+-}_{\perp i'} ) = 5 ~,~~~
{\cal E}_0 ( \psi^{++}_{\perp i}, \psi^{--}_{\perp i'} ) = 7 ~,~~~
{\cal E}_0 ( \psi^{+-}_{\perp i}, \psi^{-+}_{\perp i'} ) = 8 ~.
\end{equation}
The equations of motion for the other physical modes of gravitino are
grouped into two sets according to the $SO(4)$ chirality.  As for the
positive chirality, we have
\begin{eqnarray}
 & &  ( \Box + \frac{\mu}{4} \gamma^{123} \partial_-  ) \psi_4^+
 - \frac{3 \mu}{8} \gamma^{1234} \partial_- \psi'^+_\|
 + \frac{5\mu}{8 \sqrt{2}} \partial_- \lambda^+ = 0 ~,
  \nonumber \\
 & &  ( \Box + \frac{7 \mu}{6} \gamma^{123} \partial_- ) \psi'^+_\|
 + \frac{5 \mu}{3} \gamma^9 \partial_- \psi_4^+
 + \frac{\mu}{2\sqrt{2}} \gamma^4 \partial_- \lambda^+ = 0 ~,
\label{sg+}
\end{eqnarray}
while for the negative chirality
\begin{eqnarray}
 & & \Box \psi_4^-
 - \frac{3 \mu}{8} \gamma^{1234} \partial_- \psi'^-_\|
 + \frac{\mu}{4 \sqrt{2}} \partial_- \lambda^- = 0 ~,
  \nonumber \\
 & & ( \Box + \frac{5 \mu}{6} \gamma^{123} \partial_- ) \psi'^-_\|
 - \frac{4 \mu}{3} \gamma^9 \partial_- \psi_4^-
 - \frac{\mu}{\sqrt{2}} \gamma^4 \partial_- \lambda^- = 0 ~.
\label{sg-}
\end{eqnarray}
We note that the modes $\psi^\pm_\|$ are not dynamical because
$\gamma^I \psi_I^\pm = 0$ as can be seen from Eq.~(\ref{m-}).  Thus
the equations of motion for them have not been considered.

We see that the physical modes $\lambda^+$, $\psi^+_4$, and
$\psi'^+_\|$ are linked with each other through their equations of
motion, Eqs.~(\ref{sd+}) and (\ref{sg+}).  This is also the case for
$\lambda^-$, $\psi^-_4$, and $\psi'^-_\|$ from Eqs.~(\ref{sd-}) and
(\ref{sg-}).  Diagonalization procedure is thus required to obtain the
spectrum of normal excitation modes.  In order to do that, it is
convenient to introduce the following definitions:
\begin{equation}
\lambda^\pm \equiv - \sqrt{2} \gamma^9 \hat{\lambda}^\pm~,~~~
\psi^\pm_4 \equiv \gamma^4 \hat{\psi}^\pm_4 ~.
\label{redef}
\end{equation}
For the modes with positive $SO(4)$ chirality, we then have from
Eqs.~(\ref{sd+}) and (\ref{sg+})
\begin{eqnarray}
& & ( \Box - \frac{7\mu}{12} \gamma^{123} \partial_- )
   \hat{\lambda}^+
   + \frac{\mu}{2} \gamma^{123} \partial_- \hat{\psi}^+_4
   + \frac{\mu}{4} \gamma^{123} \partial_- \psi'^+_\| = 0 ~,
            \nonumber \\
& & ( \Box - \frac{\mu}{4} \gamma^{123} \partial_- )
   \hat{\psi}^+_4
   + \frac{5\mu}{8} \gamma^{123} \partial_- \hat{\lambda}^+
   + \frac{3\mu}{8} \gamma^{123} \partial_- \psi'^+_\| = 0 ~,
            \nonumber \\
& & ( \Box + \frac{7\mu}{6} \gamma^{123} \partial_- )
   \psi'^+_\|
   + \frac{\mu}{2} \gamma^{123} \partial_- \hat{\lambda}^+
   + \frac{5\mu}{3} \gamma^{123} \partial_- \hat{\psi}^+_4 = 0 ~.
\end{eqnarray}
The diagonalization of these equations is done by defining
\begin{eqnarray}
\chi^+_1 & \equiv &
     - \frac{1}{2} \hat{\lambda}^+
     - \frac{3}{4} \hat{\psi}^+_4 + \psi'^+_\| ~,
          \nonumber \\
\chi^+_3 & \equiv &
     \frac{3}{2} \hat{\lambda}^+
     - \frac{7}{4} \hat{\psi}^+_4 + \psi'^+_\| ~,
          \nonumber \\
\chi^+_5 & \equiv &
      \frac{1}{6} \hat{\lambda}^+
     + \frac{1}{4} \hat{\psi}^+_4 + \psi'^+_\| ~,
\end{eqnarray}
whose equations of motion are then obtained as
\begin{equation}
( \Box - \frac{\mu}{3} \gamma^{123} \partial_- )
     \chi^+_1 = 0 ~,~~~
( \Box - \mu \gamma^{123} \partial_- )
     \chi^+_3 = 0 ~,~~~
( \Box + \frac{5 \mu}{3} \gamma^{123} \partial_- )
     \chi^+_5 = 0 ~.
\end{equation}
If we split the modes in a way that respects the $SO(3)$
structure and follow the notation of Eq.~(\ref{34d}),
these equations lead to
\begin{eqnarray}
& & ( \Box - i \frac{\mu}{3} \partial_- )
         \chi^{++}_1 = 0 ~,~~~
    ( \Box + i \frac{\mu}{3} \partial_- )
         \chi^{-+}_1 = 0 ~,
\nonumber \\
& & ( \Box - i \mu \partial_- )
          \chi^{++}_3 = 0 ~,~~~
    ( \Box + i \mu \partial_- )
          \chi^{-+}_3 = 0 ~,
\nonumber \\
& & ( \Box + i \frac{5 \mu}{3} \partial_- )
     \chi^{++}_5 = 0 ~,~~~
    ( \Box - i \frac{5 \mu}{3} \partial_- )
     \chi^{-+}_5 = 0 ~,
\end{eqnarray}
from which we have the following minimal light-cone energy
spectrum for the diagonalized physical modes:
\begin{eqnarray}
& & {\cal E}_0 (\chi^{++}_1) = 7 ~,~~~
    {\cal E}_0 (\chi^{-+}_1) = 5 ~,
 \nonumber \\
& & {\cal E}_0 (\chi^{++}_3) = 9 ~,~~~
    {\cal E}_0 (\chi^{-+}_3) = 6 ~,
 \nonumber \\
& & {\cal E}_0 (\chi^{++}_5) = 1 ~,~~~
    {\cal E}_0 (\chi^{-+}_5) = 11 ~.
\end{eqnarray}

We now turn to the modes with negative $SO(4)$ chirality, that is,
$\lambda^-$, $\psi^-_4$ and $\psi'^-_\|$.  With the redefinitions,
(\ref{redef}), Eqs.~(\ref{sd-}) and (\ref{sg-}) lead to
\begin{eqnarray}
& & ( \Box - \frac{\mu}{6} \gamma^{123} \partial_- )
   \hat{\lambda}^-
   + \frac{\mu}{4} \gamma^{123} \partial_- \psi'^-_\| = 0 ~,
            \nonumber \\
& & \Box \hat{\psi}^-_4
   - \frac{\mu}{4} \gamma^{123} \partial_- \hat{\lambda}^-
   + \frac{3\mu}{8} \gamma^{123} \partial_- \psi'^-_\| = 0 ~,
            \nonumber \\
& & ( \Box + \frac{5\mu}{6} \gamma^{123} \partial_- )
   \psi'^-_\|
   + \mu \gamma^{123} \partial_- \hat{\lambda}^-
   + \frac{4\mu}{3} \gamma^{123} \partial_- \hat{\psi}^-_4 = 0 ~.
\end{eqnarray}
The diagonalized equations of motion are obtained as
\begin{equation}
\Box \chi^-_0 = 0 ~,~~~
( \Box - \frac{2 \mu}{3} \gamma^{123} \partial_- )
     \chi^-_2 = 0 ~,~~~
( \Box + \frac{4 \mu}{3} \gamma^{123} \partial_- )
     \chi^-_4 = 0 ~,
\label{fdia2}
\end{equation}
where the normal excitation modes are defined by
\begin{eqnarray}
\chi^-_0 & \equiv &
      \frac{3}{2} \hat{\lambda}^-
     - \frac{7}{4} \hat{\psi}^-_4 + \psi'^-_\| ~,
          \nonumber \\
\chi^-_2 & \equiv &
     -\frac{1}{2} \hat{\lambda}^-
     - \frac{3}{4} \hat{\psi}^-_4 + \psi'^-_\| ~,
          \nonumber \\
\chi^-_4 & \equiv &
      \frac{1}{6} \hat{\lambda}^-
     + \frac{1}{4} \hat{\psi}^-_4 + \psi'^-_\| ~.
\end{eqnarray}
After taking into account the $SO(3)$ structure with the notation of
Eq.~(\ref{34d}), the above set of equations, (\ref{fdia2}), leads us to
have
\begin{eqnarray}
& & \Box \chi^{+-}_0 = 0 ~,~~~
    \Box \chi^{--}_0 = 0 ~,
 \nonumber \\
& & ( \Box - i \frac{2 \mu}{3} \partial_- )
         \chi^{+-}_2 = 0 ~,~~~
    ( \Box + i \frac{2 \mu}{3} \partial_- )
         \chi^{--}_2 = 0 ~,~
 \nonumber \\
& & ( \Box + i \frac{4 \mu}{3} \partial_- )
         \chi^{+-}_4 = 0 ~,~~~
    ( \Box - i \frac{4 \mu}{3} \partial_- )
         \chi^{--}_4 = 0 ~.
\end{eqnarray}
According to Eq.~(\ref{lstE}), we then have the minimal light-cone
energy for the modes as
\begin{eqnarray}
& & {\cal E}_0 (\chi^{+-}_0) = 6 ~,~~~
    {\cal E}_0 (\chi^{--}_0) = 6 ~,
 \nonumber \\
& & {\cal E}_0 (\chi^{+-}_2) = 8 ~,~~~
    {\cal E}_0 (\chi^{--}_2) = 4 ~,
 \nonumber \\
& & {\cal E}_0 (\chi^{+-}_4) = 2 ~,~~~
    {\cal E}_0 (\chi^{--}_4) = 10 ~.
\end{eqnarray}

\begin{table}
\begin{center}
\begin{tabular}{|c|lr|lr|}
\hline
${\cal E}_0$
    & Bosonic mode $(N_{d.o.f.})$
    & $N_B$
    & Fermionic mode $(N_{d.o.f.})$
    & $N_F$  \\
\hline
0   & $\phi_6$(1)
    & 1
    &
    & 0 \\
1   &
    & 0
    & $\chi^{++}_5$(4)
    & 4 \\
2   & $\beta_{4i}$(3) $\beta_{4i'j'}$(3)
    & 6
    & $\chi^{+-}_4$(4)
    & 4 \\
3   & $\beta_{3i'}$(4) $\beta_{3ijk'}$(12)
    & 16
    & $\chi^{-+}_3$(4)
    & 4 \\
4   & $\phi_2$(1) $\beta_{2i}$(3) $\beta_{2i'j'}$(3)
    & 7
    & $\chi^{--}_2$(4) $\psi^{--}_{\perp i}$(8)
      $\psi^{++}_{\perp i'}$(12)
    & 24 \\
5   & $\beta_{1i'}$(4) $\beta_{1ij'}$(12)
    & 16
    & $\chi^{-+}_1$(4) $\psi^{-+}_{\perp i}$(8)
      $\psi^{+-}_{\perp i'}$(12)
    & 24 \\
6   & $\phi_0$(1) $\beta_{0i}$(3) $\beta_{0ij'k'}$(18)
      $h^\perp_{ij}$(5) $h^\perp_{i'j'}$(9)
    & 36
    & $\chi^{+-}_0$(4) $\chi^{--}_0$(4)
    & 8 \\
7   & $\bar{\beta}_{1i'}$(4) $\bar{\beta}_{1ij'}$(12)
    & 16
    & $\chi^{++}_1$(4) $\psi^{++}_{\perp i}$(8)
      $\psi^{--}_{\perp i'}$(12)
    & 24 \\
8   & $\bar{\phi}_2$(1) $\bar{\beta}_{2i}$(3) $\bar{\beta}_{2i'j'}$(3)
    & 7
    & $\chi^{+-}_2$(4) $\psi^{+-}_{\perp i}$(8)
      $\psi^{-+}_{\perp i'}$(12)
    & 24 \\
9   & $\bar{\beta}_{3i'}$(4) $\bar{\beta}_{3ijk'}$(12)
    & 16
    & $\chi^{++}_3$(4)
    & 4 \\
10  & $\bar{\beta}_{4i}$(3) $\bar{\beta}_{4i'j'}$(3)
    & 6
    & $\chi^{--}_4$(4)
    & 4 \\
11  &
    & 0
    & $\chi^{-+}_5$(4)
    & 4 \\
12  & $\bar{\phi}_6$(1)
    & 1
    &
    & 0 \\
\hline
    &  & 128 &  & 128 \\
\hline
\end{tabular}
\end{center}
\caption{Type IIA supergravity excitation modes.  $N_{d.o.f}$ means
  the number of degrees of freedom.  $N_B$ ($N_F$) is the number of
  bosonic (fermionic) degrees of freedom at each light-cone energy.}
\label{sugra-states}
\end{table}

\section{Conclusion and discussion}
\label{concl}

We have obtained the Type IIA supergravity excitation modes and their
spectrum in the pp-wave background, which are summarized in table
\ref{sugra-states}.  The supergravity modes have been arranged such
that they respect the $SO(3) \times SO(4)$ symmetry structure of the
background.  We see that there is mismatch between the bosonic and
fermionic degrees of freedom at each light-cone energy, basically due
to the fact that the supersymmetry preserved by the pp-wave background
is time-dependent. (More precisely, 16 among 24 supersymmetries depend
on the light-cone time $x^+$ \cite{hyu074}.)

The result we have obtained shows how the low-lying string states
listed in table \ref{string-states} correspond to the supergravity
modes of table \ref{sugra-states}.  This implies that we can associate
the vertex operator for a certain low-lying string state to a definite
supergravity excitation mode in the pp-wave background.  It may be
expected that the study of string amplitudes with such vertex
operators gives useful insight into the structure of the M theory in
the maximally supersymmetric eleven-dimensional pp-wave background.

Concerning the eleven dimensional origin of the ten dimensional
supergravity excitation spectrum, it would be interesting to uncover
the structure of the spectrum obtained in this paper.  The eleven
dimensional perturbative spectrum from the Matrix model in the eleven
dimensional pp-wave background has intriguing features such as the
protected multiplet and the indication of the presence of the
transverse five-brane \cite{das050,mal139}.  Since the Type IIA
supergravity inherits most of its features through the dimensional
reduction, the structure in eleven dimensions would be encoded in the
spectrum of ten dimensional supergravity.  Thus the eleven dimensional
perturbative spectrum would lead us to have the deeper understanding
of the ten dimensional spectrum.  What we would get from the physics
related to the transverse five-brane is particularly interesting.  We
hope to return to this issue in the near future \cite{kwo}.

\section*{Acknowledgments}
We would like to thank Seungjoon Hyun and Jeong-Hyuck Park for helpful
discussions.  This work of O.K. is the result of research activities
(Astrophysical Research Center for the Structure and Evolution of the
Cosmos (ARCSEC) and the Basic Research Program, R01-2000-000-00021-0)
supported by Korea Science $\&$ Engineering Foundation.  The work of
H.S. was supported by Korea Research Foundation Grant
KRF-2001-015-DP0082.

\end{document}